\documentclass[prd,onecolumn,aps,amsfonts,
floats,tightenlines,floatfix,superscriptaddress]{revtex4}
\usepackage[english]{babel}
\usepackage{epsfig}
\usepackage{psfrag}
\usepackage{graphicx}
\usepackage{amsmath}
\usepackage{amssymb}
\usepackage{dsfont}
\usepackage{slashed}
\newcommand{\intq}{\int\!\!\frac{d^4q}{(2 \pi)^4}}

\newcommand{\pslash}{\slashed{p}}

\newcommand{\phslash}{\slashed{\hat{p}}}

\newcommand{\wpslash}{\slashed{\omega}_{p}}
\newcommand{\pvslash}{\slashed{\vec{p}}}

\begin{document}

\date{\today}

\title{On the unlocking of color and flavor in color-superconducting quark
  matter}

\author{D.~Nickel}
\affiliation{Institute for Nuclear Physics, Technical University Darmstadt, 
  Schlo{\ss}gartenstra{\ss}e 9, D-64289 Darmstadt, Germany}
\affiliation{Institute of Physics, University of Graz,
  Universit{\"a}tsplatz 5, A-8010 Graz, Austria}
\author{R. Alkofer}
\affiliation{Institute of Physics, University of Graz,
  Universit{\"a}tsplatz 5, A-8010 Graz, Austria}
\author{J.~Wambach}
\affiliation{Institute for Nuclear Physics, Technical University Darmstadt, 
  Schlo{\ss}gartenstra{\ss}e 9, D-64289 Darmstadt, Germany}
\affiliation{Gesellschaft f{\"u}r Schwerionenforschung mbH, Planckstra{\ss}e
  1, D-64291 Darmstadt, Germany}

\begin{abstract}
The role of the strange quark mass for the phase structure of QCD at
non-vanishing densities is studied by employing a recently developed
self-consistent truncation scheme for the Dyson-Schwinger equations of the
quark propagators in Landau gauge.
Hereby the medium modification of the effective quark interaction by
the polarization of gluons is implemented. Taking into account this effect
results in significantly smaller dynamical quark masses at the Fermi surface. 
Due to this reduction the color-flavor locked phase is always the preferred
color-superconducting phase at zero temperature and for a realistic strange
quark mass.
\end{abstract}

\maketitle

\section{Introduction}
The possible occurrence of different phases in Quantum Chromo Dynamics
(QCD) at sufficiently high  densities and small temperatures has attracted a
lot of interest in recent  years (for corresponding reviews see {\it e.g.\/}
refs.\ \cite{reviews,Rischke:2003mt}). Within models, and depending on various
parameters and constraints, a surprisingly rich structure of different
color-superconducting phases has been studied. The pairing pattern of
quarks  strongly depends on the relative position of the Fermi surfaces of the
different quark species. Therefore the dynamically generated constituent
masses 
as well as the effective chemical potentials are the relevant quantities for 
a study of color-flavor unlocking.

The color-flavor locked (CFL) phase is expected to be the ground state for a
three flavor color-superconductor at asymptotically large
densities~\cite{Alford:1998mk}. This raises the question of what happens to
this state when the density is lowered to more realistic values. Since the CFL
phase is neutral for degenerate quarks this depends
mainly on the mass splitting between the strange quark and the two light (up
and down) quarks. The unlocking of color and flavor has been investigated
within models of the Nambu \& Jona-Lasinio (NJL) type by varying the strange
quark mass~\cite{Alford:1999pa,Schafer:1999pb}. It has been
concluded that at some densities the CFL phase is disfavored compared to the
2-flavor color superconductor (2SC) phase in a self-consistent
treatment~\cite{Buballa:2001gj}.
The region of the  2SC phase possesses a rich (sub-)structure if furthermore
neutrality conditions and $\beta$-equilibrium are taken into
account~\cite{Steiner:2002gx}. The CFL phase and the boundary for unlocking
are, however, only mildly affected.

These results, obtained in NJL-type models, are rather sensitive to model
parameters. For more reliable statements a non-perturbative approach,
which is directly based on the QCD degrees of freedom and recovers the
correct vacuum results as well as the proper weak coupling behavior at
asymptotically large densities, is needed.

For the description of QCD vacuum properties, truncated Dyson-Schwinger
equations (DSEs) in Landau gauge have been successfully applied (for
corresponding reviews see refs.\ \cite{reviews2}). They naturally describe 
gluon \cite{Alkofer:2003jj} as well as quark \cite{Alkofer:2006gz} confinement.
They have furthermore been extended to non-vanishing chemical potentials with
the correct behavior in the weak coupling limit~\cite{Nickel:2006vf}. Therefore
this method will be employed here for the investigation of the color-flavor
unlocking. Within this framework, we present self-consistent solutions of  the
quark propagators in the 2SC and CFL phase for realistic values of the
strange-quark current mass. We include energy- and momentum-dependent dressing
functions for the most general parameterization in both Dirac and color-flavor
space. For the reasons stated above, we neglect for the time being the
neutrality conditions in the present investigation of color-flavor
unlocking. In contrast to studies employing  NJL-type models, we have also
implemented a medium modification of the effective quark interaction taking
into account the medium contribution to the polarization tensor which enters
the gluon propagator.  It has to be noted that a proper description of the
weak coupling limit requires the inclusion of this effect. The vacuum part of
the effective quark interaction is taken either from a solution of coupled
gluon, ghost and quark propagator DSEs or from a fit to lattice data for the
gluon and the quark propagators~\cite{Fischer:2003rp,Bhagwat:2003vw}.

In the chiral limit and at moderate values of the quark density, the gap
functions at the Fermi surface in such a calculation acquire sizeable 
values~\cite{Nickel:2006vf}, comparable to those obtained
within  NJL-type models. These results are found to be rather insensitive to
the effective quark interaction, at least within the range of uncertainty of
the effective coupling.  The dynamically generated mass functions, on the
other
hand, turn out to be significantly smaller than in the vacuum. This is due to
the medium modification of the interaction (which is usually neglected in
NJL-type investigations). This effect of smaller dynamical quark masses implies
that the Fermi surfaces of the quarks are closer to each other than expected
and eventually leads to  the most important result presented here: the CFL
phase is, at zero temperature and realistic values for the strange quark
current mass, always energetically preferred compared to the 2SC phase.

This paper is organized as follows: In section~\ref{setup} we briefly present 
the employed theoretical framework, leading to a fully self-consistent
treatment of the CFL phase for a range of strange quark masses. A physical
interpretation of the parameterization used is provided. In
section~\ref{results} we discuss the numerical results including those for
Fermi momenta, mass  and gap functions, etc.. Most importantly,  we provide a
prediction for the critical value for the strange quark current mass as a
function of the chemical potential. Finally, we summarize and conclude in
section~\ref{conclusions}. The  color-flavor basis employed is given in the
Appendix.

\section{The Landau-gauge quark propagator at non-vanishing chemical
potential}
\label{setup}
\subsection{The truncated quark DSE at non-vanishing chemical potential}
In ref.\ \cite{Nickel:2006vf}, a closed truncated DSE for the Landau-gauge 
quark propagator at non-vanishing chemical potential has been derived and 
solved in the chiral limit. In this work we are following the scheme and
notations presented there. In the Nambu-Gor'kov basis (see ref.\
\cite{Rischke:2003mt} for a corresponding
review) the normal quark DSE is coupled to an equation for the
gap functions. To describe their structure we define
\begin{eqnarray}
\mathcal{S}_{0}(p)&=&
\left(
  \begin{array}{cc}
    S_{0}^{+}(p) & 0\\
    0 & S_{0}^{-}(p) = -CS_{0}^{+}(-p)^{T}C
  \end{array}
\right),\\
\mathcal{S}(p)&=&
\left(
  \begin{array}{cc}
    S^{+}(p)& T^{-}(p) = \gamma_{4}T^{+}(p)^{\dagger}\gamma_{4}\\
    T^{+}(p)& S^{-}(p) = -CS^{+}(-p)^{T}C
  \end{array}
\right),\\
\Sigma(p)&=&
\left(
  \begin{array}{cc}
    \Sigma^{+}(p)&\Phi^{-}(p)\\
    \Phi^{+}(p)&\Sigma^{-}(p)
  \end{array}
\right),
\end{eqnarray}
with
$\mathcal{S}_{0}$ being the bare and $\mathcal{S}$ the full Nambu-Gor'kov
propagator. $\Sigma$ is the Nambu-Gor'kov self-energy. $S_{0}^{+}$ is diagonal
in color-flavor space and equal to
$S_{0,f}^{+}(p)^{-1}= -i(p_{4}+i\mu)\gamma_{4}-i\vec{p}\cdot\vec{\gamma}+m_{0,f}$
for a quark flavor $f$. As the Euclidean action is real we furthermore have to
impose
$\Phi^{-}(p)   = \gamma_{4}\Phi^{+}(p)^{\dagger}  \gamma_{4}$.
Introducing appropriate renormalization constants, the DSE for 
the quark propagator
\begin{eqnarray}
\label{qDSE}
  \mathcal{S}^{-1}(p) &=& Z_{2}\mathcal{S}_{0}^{-1}(p)+ Z_{1F}\Sigma(p) 
\end{eqnarray}
explicitly leads to
\begin{eqnarray}
  \label{fullT}
  T^{\pm} &=&
  -Z_{1F}\left(Z_{2}{S^{\mp}_{0}}^{-1}+Z_{1F}\Sigma^{\mp}\right)^{-1}
  \Phi^{\pm}S^{\pm},\\
  \label{fullS}
  {S^{\pm}}^{-1} &=& \hphantom{-}Z_{2}{S^{\pm}_{0}}^{-1}+Z_{1F}\Sigma^{\pm}-
  Z_{1F}^{2}\Phi^{\mp}\left(Z_{2}{S^{\mp}_{0}}^{-1}+Z_{1F}\Sigma^{\mp}
  \right)^{-1}\Phi^{\pm}.
\end{eqnarray}
Employing the truncation of ref.\ \cite{Nickel:2006vf}, the equations for the 
self-energy and the gap function then read
\begin{eqnarray}
  \Sigma^{+}(p) &=& \hphantom{-}\frac{Z_{2}^{2}}{Z_{1F}} \pi \intq
  \gamma_{\mu} \lambda_{a} S^{+}(q) \gamma_{\nu} \lambda_{a}
  \left(
    \frac{\alpha_{s}(k^{2}) P^{T}_{\mu\nu}}{k^{2}+G(k)}+
    \frac{\alpha_{s}(k^{2}) P^{L}_{\mu\nu}}{k^{2}+F(k)}
  \right),
  \\
  \Phi^{+}(p) &=& -\frac{Z_{2}^{2}}{Z_{1F}} \pi \intq
  \gamma_{\mu} \lambda_{a}^{T} T^{+}(q) \gamma_{\nu} \lambda_{a}
  \left(
    \frac{\alpha_{s}(k^{2}) P^{T}_{\mu\nu}}{k^{2}+G(k)}+
    \frac{\alpha_{s}(k^{2}) P^{L}_{\mu\nu}}{k^{2}+F(k)}
  \right),
\end{eqnarray} 
where $k=p-q$. Here projectors transverse and longitudinal in the medium have
been introduced. The functions $G$ and $F$ describe the corresponding
medium modifications of the gluon propagator, see Eq.\ (21)
of ref.\ \cite{Nickel:2006vf} and are calculated, once the coupling is given.
Therefore the only input for the quark DSE are the running coupling 
$\alpha_{s}(k^{2})$ (for its choice see below) and the renormalized 
quark current masses.

Before proceeding we note that this quark DSE can also be derived from the
thermodynamically consistent effective action corresponding to the truncation
scheme used~\cite{Nickel:2006vf}. Evaluating this action at the stationary
point one obtains:
\begin{eqnarray}
  \label{CJTeqn}
  \Gamma[\mathcal{S}] &=& 
  -\frac{1}{2}\mathrm{Tr}_{p,D,c,f,NG}\mathrm{Ln}\mathcal{S}^{-1}+
  \frac{1}{4}\mathrm{Tr}_{p,D,c,f,NG}
  \left(1-Z_{2}\mathcal{S}_{0}^{-1}\mathcal{S}\right)+
  const.
  \nonumber\\&=&
  -\frac{1}{2}\mathrm{Tr}_{p,D,c,f}\mathrm{Ln}\left(
    {S^{+}}^{-1}
    \left(Z_{2}S_{0}^{-\,-1}+Z_{1F}\Sigma^{-}\right)
  \right)+
  \nonumber\\&&
  +
  \frac{1}{4}\mathrm{Tr}_{p,D,c,f}
  \left(
    2-Z_{2}S^{+}S_{0}^{+\,-1}-Z_{2}S^{-}S_{0}^{-\,-1}
  \right)+
  const.
\end{eqnarray}
Later on we will use this expression to determine {\it e.g.\/} 
the pressure difference  between different phases at a given chemical potential.

Similar to our previous investigation in the chiral limit~\cite{Nickel:2006vf}, 
we employ two different couplings. The running coupling determined in DSE
studies of the Yang-Mills sector~\cite{Fischer:2003rp}, which is labeled in
the following as $\alpha_{I}(k^{2})$, will serve as a lower bound. As is
detailed in ref.\ \cite{Fischer:2003rp} it underestimates chiral symmetry
breaking
significantly in the abelian approximation, which on the other hand is an
appropriate choice if the ground state does not break chiral symmetry
dynamically or explicitly, see ref.\ \cite{Alkofer:2006gz} and references
therein. The running coupling extracted from lattice QCD data for the
quark and gluon propagators~\cite{Bhagwat:2003vw} will serve as an upper
bound, since the medium will probably weaken the interaction. It is labeled
$\alpha_{II}(k^{2})$. In order to vary the renormalization point in the latter
version of the coupling, we use the multiplicative renormalizibility of the
quark DSE. Note also that the vacuum quark propagator, determined within such
a scheme~\cite{Fischer:2005nf} is in excellent agreement with the
corresponding lattice data when the finite volume of lattice calculations is
accounted for.

As stated above, taking into account the medium polarization, Debye screening
and Landau damping are included. Both, screening and damping of the
interaction increase for increasing interaction strength. Therefore the
generated gap and  mass (normal self-energy) functions turn out to be much less
sensitive against variation of $\alpha_{s}(k^{2})$ than the dynamical mass
function in the  chirally broken vacuum. Due to this fortunate instance the
presented approach has considerable predictive power despite the uncertainty
in the effective low-energy quark interaction at non-vanishing densities.

The renormalization constants are determined in the (chirally broken) vacuum.
Due to the employed vertex construction, the quark-gluon vertex renormalization
constant, $Z_{1F}$, cancels in the resulting renormalized equations.
For each flavor, we determine the quark wave-function renormalization
constant, $Z_{2}$, and the renormalization constant $Z_{m}$, relating the
unrenormalized quark mass $m_{0,q}(\Lambda^{2})$ at an ultraviolet 
cutoff $\Lambda$ to the renormalized mass $m_{q}(\nu)$ via
\begin{eqnarray}
  m_{0,q}(\Lambda^{2}) &=& Z_{m}(\nu^{2},\Lambda^{2})m_{q}(\nu),
\end{eqnarray}
by requiring
\begin{eqnarray}
  \left. S^{+}_{q}(p)\right|_{p^{2}=\nu^{2}} &=& -i\pslash+m_{q}(\nu)
\end{eqnarray}
at a renormalization scale $\nu$.
This corresponds to a momentum-subtraction ($MOM$) scheme, which results in
somewhat smaller numerical values for the quark current masses at a given
renormalization scale (usually taken to be $2\,\mathrm{GeV}$). We simply ignore
here the
difference between $\overline{MS}$ and $MOM$ masses because the effect is of
the order of ten percent (when calculated within perturbation theory) and thus
much smaller than the uncertainty in the value the current masses~\footnote{
In addition, it is worth mentioning that solutions of the Bethe-Salpeter
equation within the DSE approach in $MOM$ scheme favor small values for the
physical strange quark-current mass ~\cite{Fischer:2005en}, a fact which is
definitely related to the difference in renormalization schemes.}.

It turns out that, as expected, the mass dependence of the quark wave function
renormalization constant, $Z_{2}$, is negligible as long as the renormalization
scale is much larger than the mass. Therefore, we simply drop this dependence
and $Z_{2}$ is determined once and for all in the chiral limit.  To keep the
number of parameters as small as possible, we work in the chiral isospin limit 
({\it i.e.\/} we set the up and down current quark masses to zero) and vary
only the strange quark current mass.

In the following we will restrict to isotropic phases. In order to solve the
DSE of the quark propagator it is advantageous to consider their color-flavor
structure first. To get a self-consistent solution we choose suitable sets of
matrices $\{P_{i}\}$ and $\{M_{i}\}$ in color-flavor space, such that
\begin{eqnarray}
  \Sigma^{+}(p)
  &=&
  \frac{Z_{2}}{Z_{1F}}
  \sum_{i} \Sigma^{+}_{i}(p) P_{i},\\
  \Phi^{+}(p)
  &=&
  \frac{Z_{2}}{Z_{1F}}
  \sum_{i} \phi^{+}_{i}(p) M_{i},
\end{eqnarray}
where we have introduced the renormalization-point independent component
functions $\Sigma^{+}_{i}(p)$ and $\phi^{+}_{i}(p)$, which are matrix-valued
in  Dirac space. Full self-consistency is guaranteed in case a basis of all
allowed matrices is considered. The dimensionality of this basis in a given phase
depends on the residual symmetry in color-flavor space. For the CFL phase
this will be detailed below.

The Dirac structure of the self-energies in an even-parity phase can be
para\-meterized by~\cite{Pisarski:1999av}
\begin{eqnarray}
  \Sigma^{+}_{i}(p) &=&
  -i\phslash\,\Sigma^{+}_{A,i}(p)-i\wpslash\,\Sigma^{+}_{C,i}(p)
  +\Sigma^{+}_{B,i}(p)-i\gamma_{4}\phslash\,\Sigma^{+}_{D,i}(p), 
  \\
  \label{phi}
  \phi^{+}_{i}(p) &=& 
  \left(
    \gamma_{4}\phslash\,\phi^{+}_{A,i}(p)+\gamma_{4}\,\phi^{+}_{B,i}(p)
    +\phi^{+}_{C,i}(p)+\phslash\,\phi^{+}_{D,i}(p)
  \right)\gamma_{5},
\end{eqnarray}
where $\hat{p}=\vec{p}/\vert \vec{p}\vert$, $\phslash =\hat{p}\cdot\vec{\gamma}$, 
$\wpslash = \omega_{p}\gamma_{4}$ and $\omega_{p}=ip_{4}+\mu$.
We already note that $\Sigma^{+}_{D,i}(p)$ turns out to be negligibly small. 
This is to be expected since it has to vanish due to time reversal symmetry at 
vanishing temperatures in color-flavor symmetric channels. The gap functions
$\phi_{B}$ and $\phi_{D}$, which vanish in the chiral limit will be 
discussed in subsection \ref{gapinter} where we provide an interpretation 
in a simplified setting.

The system of equations for the self-energies is symmetric under the
transformations
\begin{eqnarray}
  \Sigma^{+}_{A/B/C,i}(p_{4},\vert\vec{p}\vert) &\rightarrow&
  \phantom{-}\Sigma^{+}_{A/B/C,i}(-p_{4},\vert\vec{p}\vert)^{*},
  \\
  \Sigma^{+}_{D,i}(p_{4},\vert\vec{p}\vert) &\rightarrow&
  -\Sigma^{+}_{D,i}(-p_{4},\vert\vec{p}\vert)^{*},
\end{eqnarray}
which we find to be unbroken in the self-consistent scheme presented
here. Therefore, the determinant of the quark propagator is positive for
$p_{4}=0$, and the Luttinger theorem is applicable for determining the
density~\cite{Nickel:2006vf}.

For the anomalous propagator an inhomogeneous part is missing in
Eq.(\ref{fullT}). Therefore we can choose a global phase for the gap
functions. This is done such that the equations are invariant under
\begin{eqnarray}
  \phi^{+}_{A/D,i}(p_{4},\vert\vec{p}\vert) &\rightarrow&
  \phantom{-}\phi^{+}_{A/D,i}(-p_{4},\vert\vec{p}\vert)^{*}
  \\
  \phi^{+}_{B/C,i}(p_{4},\vert\vec{p}\vert) &\rightarrow&
  -\phi^{+}_{B/C,i}(-p_{4},\vert\vec{p}\vert)^{*},
\end{eqnarray}
and therefore $\phi_{A}$ and $\phi_{D}$ are real, and $\phi_{B}$ and
$\phi_{C}$ purely imaginary, for $p_{4}=0$.

Neglecting $\Sigma^{+}_{D,i}$, the Dirac structure of the inverse normal
propagator can be written as
\begin{eqnarray}
  \label{inverseSp}
  {S^{+}_{i}}^{-1}(p) &=& Z_{2}{S^{+}_{0,i}}^{-1}+Z_{1F}\Sigma^{+}_{i}
  = -i\pvslash A_{i}(p)-i\wpslash C_{i}(p)+B_{i}(p),
\end{eqnarray}
which allows to define the mass function for the pairing quasiparticles
for a given color-flavor channel $q$ as
\begin{eqnarray}
  M_{q}(p) &=& \frac{m_{0,q}+\Sigma^{+}_{B,q}(p)}{1+\Sigma^{+}_{C,q}(p)}.
\end{eqnarray}
In the following we will mostly present results for mass functions at 
$p_{4}=0$. To ease notations we will use $M_{q}(\vert\vec{p}\vert) =
M_{q}(p_{4}=0,\vert\vec{p}\vert)$ except when stated explicitly otherwise.

\subsection{The parameterization of the CFL phase}
\label{CFLdef}
As its name already expresses, the CFL phase is defined via its symmetry
pattern in color-flavor space. For three degenerate quark flavors, one
considers the $SU(3)_{c+V}$ symmetry generated by $\tau_{a}-\lambda_{a}^{T}$,
with $a=1,\dots ,8$ and $\tau_{a}$, $\lambda_{a}$ being the Gell-Mann matrices
in flavor and color space, respectively~\cite{Alford:1998mk}. The quarks
therefore are in a $\pmb{3}\otimes\pmb{\bar{3}}=\pmb{1}\oplus
\pmb{8}$ representation of this symmetry. In the case of only two degenerate
quarks and a strange quark, the symmetry is broken down to
$SU(2)_{c+V}\otimes U(1)_{c+V}$, which is generated by
$\tau_{a}-\lambda_{a}^{T}$ with $a=1,2,3$ and $8$, {\it i.e.\/} 
the quarks form a
$\pmb{1}\oplus\pmb{1}\oplus\pmb{2}\oplus\pmb{2}\oplus\pmb{3}$ representation.

For the Nambu-Gor'kov propagator to be invariant under this transformation, we
need to require
\begin{eqnarray}
  \label{Urequire}
  U^\dagger S^{+}(p)U = S^{+}(p),
  \quad 
  U^{T}T^{+}(p)U = T^{+}(p),
\end{eqnarray}
for $U\in SU(2)_{c+V}\otimes U(1)_{c+V}$. The matrices $\{P_{i}\}$ and
$\{M_{i}\}$, needed for a self-consistent description are then those 
given explicitely in the Appendix.

In section~\ref{results} we will also present results for the self-energies
evaluated at different energies and momenta, in particular at the Fermi
surface. In contrast to the chiral limit, where the matrices $\{P_{i}\}$ can
be chosen as constant projectors onto irreducible representations of the
residual
symmetry, the situation is more complex here. In principle, every dressing
function of the normal propagator and self-energy can be decomposed into
irreducible projectors in color-flavor space:
\begin{eqnarray}
  \label{SCFL}
  F &=&
  F_{\pmb{1}}P_{\pmb{1}}+F_{\pmb{1}'}P_{\pmb{1}'}+
  F_{\pmb{2}}P_{\pmb{2}}+F_{\pmb{\bar{2}}}P_{\pmb{\bar{2}}}+
  F_{\pmb{3}}P_{\pmb{3}},
\end{eqnarray}
where, in particular, $P_{\pmb{2}}=P_{7}$, $P_{\pmb{\bar{2}}}=P_{8}$ and
$P_{\pmb{3}}=P_{1}-\frac{1}{2}P_{2}$. However, for every dressing function,
the singlets are not protected against mixing and therefore their projectors
$P_{\pmb{1}}$ and $P_{\pmb{1}'}$ are in general energy and momentum
dependent.
We therefore find six different Fermi momenta, which are defined as sign
changes of the determinant of the propagator,
$\det(\mathcal{S}(p_{4}=0,\vert\vec{p}\vert))$,
at $p_{4}=0$ and connected to the density by the Luttinger theorem. Three of
them correspond to the upper $3\times 3$ block-matrix in the ansatz employed 
(see Appendix) and describe the mixing of the two singlet channels as well as
the triplet channel. After ordering, these are denoted by 
$p_{F,\pmb{1}_{1}}$, $p_{F,\pmb{1}_{2}}$ and $p_{F,\pmb{1}_{3}}$.
In the lower $6\times 6$ diagonal block-matrix, we find $p_{F,\pmb{2}}$,
$p_{F,\pmb{\bar{2}}}$ and $p_{F,\pmb{3}}$ corresponding to $P_{7}$, $P_{8}$
and $P_{6}$, respectively.

For the dressing functions of the anomal propagators and the gap functions
the situation is similar. With
$M=\lambda_{2}\otimes\tau_{2}+\lambda_{5}\otimes\tau_{5}+
\lambda_{7}\otimes\tau_{7}$
being invertible and fulfilling $U^{T}MU=M$, we define $M_{i}=MP_{i}$.
Every dressing function can then be written as
\begin{eqnarray}
  \label{TCFL}
  G &=&
  G_{\pmb{1}}M_{\pmb{1}}+G_{\pmb{1}'}M_{\pmb{1}'}+
  G_{\pmb{2}}M_{\pmb{2}}+G_{\pmb{\bar{2}}}M_{\pmb{\bar{2}}}+
  G_{\pmb{3}}M_{\pmb{3}},
\end{eqnarray}
where $M_{\pmb{2}}=M_{7}$, $M_{\pmb{\bar{2}}}=M_{8}$ and
$M_{\pmb{3}}=M_{1}-\frac{1}{2}M_{2}$. Furthermore $M_{\pmb{1}}$ and
$M_{\pmb{1'}}$ are again in general energy and momentum dependent due to the
possible mixing between the singlets. 

\subsection{The gap functions $\phi_{B}$ and $\phi_{D}$}
\label{gapinter}

Having introduced the most general Dirac structure for the gap functions in an
even-parity phase (see Eq.(\ref{phi})) their interpretation and in particular
their relation to the energy gap in the excitation spectrum is of interest. To
this end one has to determine the dispersion relations as given by the poles of
the propagators. This is a solvable task but leads already for degenerate
quarks and constant dressing functions to very involved and intricate
expressions. A physically motivated approximation with a simple interpretation
of the gap function will be introduced in this subsection and will serve as an
illustration. For simplicity, we consider the pairing of only two different
quasiparticles $a$ and $b$ with inverse propagators in Dirac space, given by
${S_{a/b}^{+}}^{-1}=Z_{2}{S_{0,a/b}^{+}}^{-1}+Z_{2}\Sigma_{a/b}^{+}$ for
vanishing gap functions. This is in particular the case for the three lower
$2\times 2$ block-matrices in the ansatz presented in the Appendix.

The gapped propagator $S_{c}^{+}$ in one channel, see Eq.\ (\ref{fullS}), is then
typically given by
\begin{eqnarray}
  \label{phiinterfullS}
  {S^{+}_{c}}^{-1} &=& {S_{a}^{+}}^{-1}-
  \phi^{-}{S^{-}_{b}}\phi^{+},
\end{eqnarray}
with $\phi^{+}$ of the form in Eq.\ (\ref{phi}). The propagators, being of the
form in Eq.\ (\ref{inverseSp}), can be expressed via the energy-projectors for
$i=a,b,c$
\begin{eqnarray}
  {S^{+}_{i}}^{-1} &=&
  \sum_{e=\pm}\left(
    -i\omega C_{i} -e E_{i}
  \right)\gamma_{4}\Lambda^{e}_{i},
\end{eqnarray}
where
\begin{eqnarray}
  \Lambda^{\pm}_{i} &=&
  \frac{1}{2}\left(1\pm
  \left(i\beta_{i}\gamma_{4}\phslash+\alpha_{i}\gamma_{4}\right)\right), 
\end{eqnarray}
and $E_{i}=\sqrt{\vec{p}^{2}A_{i}^{2}+B_{i}^{2}}$, 
$\alpha_{i}={B_{i}}/{E_{i}}$ and
$\beta_{i}={\vert\vec{p}\vert A_{i}}/{E_{i}}$.
Neglecting the $p_{4}$-dependence in the dressing-functions and absorbing
$C_{a}$ and $C_{b}$ by rescaling $E_{a}$, $E_{b}$ and $\phi_{i}$ 
correspondingly~\footnote{
In principle, we also need to rescale $\mu$, formally leading to two chemical
potentials. This introduces no further complication in principle but is
neglected for the sake of simplicity.}, we assume that no pairing between
quasiparticles and anti-quasiparticles takes place. This reduces Eq.\
(\ref{phiinterfullS}) to
\begin{eqnarray}
  \left(-i\omega C_{c}-E_{c}\right)\gamma_{4}\Lambda^{+}_{c}
  &\approx&
  \left(-i\omega-E_{a}\right)\gamma_{4}\Lambda^{+}_{a}
  -
  {\phi^{-}}
  \frac{1}{-i\omega^{*}+E_{b}}
  \Lambda^{-}_{b}\gamma_{4}
  {\phi^{+}},
\end{eqnarray}
which is exact in the chiral limit. Note, that it can also be justified, at
least in leading order, in a ${\phi^{2}}/{\mu^{2}}$ expansion. Evaluating the
roots of the right-hand side, we find the dispersion relation
\begin{eqnarray}
  -ip_{4} &=&
  \frac{E_{a}-E_{b}}{2} \pm
  \bigg(
  \left(\frac{E_{a}+E_{b}}{2}-\mu\right)^{2}+
  \phi_{A}^2-\phi_{B}^2-\phi_{C}^2+\phi_{D}^2
  \nonumber\\&&
  +2i\phi_{C}\left(\beta_{2}\phi_{A}-\alpha_{2}i\phi_{B}\right)
  -2\phi_{D}\left(\alpha_{2}\phi_{A}+\beta_{2}i\phi_{B}\right)
  \bigg)^{\frac{1}{2}},
\end{eqnarray}
where we have made use of the fact that $\phi_{A}$ and $\phi_{D}$ are real and 
$\phi_{B}$ and $\phi_{C}$ are imaginary at $p_{4}=0$ (see the discussion in
the last subsection). Noting that $\alpha_{i}^{2}+\beta_{i}^{2}=1$, the above
expression suggests to introduce gap functions, which are transformed
accordingly:
\begin{eqnarray}
\tilde{\phi}_{A}=\beta_{2}\phi_{A}-\alpha_{2}i\phi_{B}, \qquad
i\tilde{\phi}_{B}=\alpha_{2}\phi_{A}+\beta_{2}i\phi_{B}.
\end{eqnarray}
This transformation depends on the mass parameter of the quasiparticle, which
makes evident that the existence of the functions $\phi_{B}$ and $\phi_{D}$
allows different energy gaps for quasiparticles with different 
masses. In terms of the `rotated` gap functions $\tilde{\phi}_{A}$ and
$\tilde{\phi}_{B}$ we obtain
\begin{eqnarray}
  -ip_{4} &=&
  \frac{E_{a}-E_{b}}{2} \pm
  \big(
  (\frac{E_{a}+E_{b}}{2}-\mu)^{2}+
  (\tilde{\phi}_{A}+i\phi_{C})^{2}+
  (\phi_{D}-i\tilde{\phi}_{B})^{2}
  \big)^{\frac{1}{2}}.
\end{eqnarray}
In analogy to the situation in the chiral limit, we arrive now at a simple
interpretation: One has a chirally symmetric pairing via
$\tilde{\phi}_{A}+i\phi_{C}$ and a chirality breaking pairing through
$\phi_{D}-i\tilde{\phi}_{B}$. For anti-quasiparticle pairing
$\tilde{\phi}_{A}$ and $\tilde{\phi}_{B}$ simply change sign.
In the chiral limit we then find, as expected, $\phi^{\pm}=\phi_{A}\pm
i\phi_{C}$ and $\phi_{B}=\phi_{D}=0$.

\section{Results}
\label{results}
As described above, the quark propagator is highly non-trivial. The Dirac
structure of the self-energies and gap functions is given by four
self-consistently determined functions, respectively, which are in addition
functions of $\vert\vec{p}\vert$ and $p_{4}$. In accordance with the aim of
this paper, namely to investigate the phenomenological importance of the
strange quark current mass, we mainly restrict ourself to the presentation of
results at $p_{4}=0$ and some values of $\vert\vec{p}\vert$. This will also
turn out to be sufficient to demonstrate the important differences of our
results as compared to corresponding ones obtained in NJL-type models. As
already mentioned, we include self-energy effects, which have been analyzed 
so far only in the weak coupling regime~\cite{Wang:2001aq}. Note that in the
chiral limit the dependence on $\vert\vec{p}\vert$ has already been discussed
in ref.\ \cite{Nickel:2006vf} and the role of a non-trivial $p_{4}$-dependence
in ref.\ \cite{Nickel:2006mm}.

The following results will be presented for the couplings discussed above,
$\alpha_{I}(k^{2})$ and $\alpha_{II}(k^{2})$.  As explained, we consider them
as the limiting cases, which are allowed by the uncertainty within
investigations of infrared QCD. We will see that gap functions and Fermi
momenta are quite insensitive to the coupling used.

\subsection{Fermi momenta}
\label{fermipf}
\begin{figure}[h!]
  \includegraphics[width=8cm]{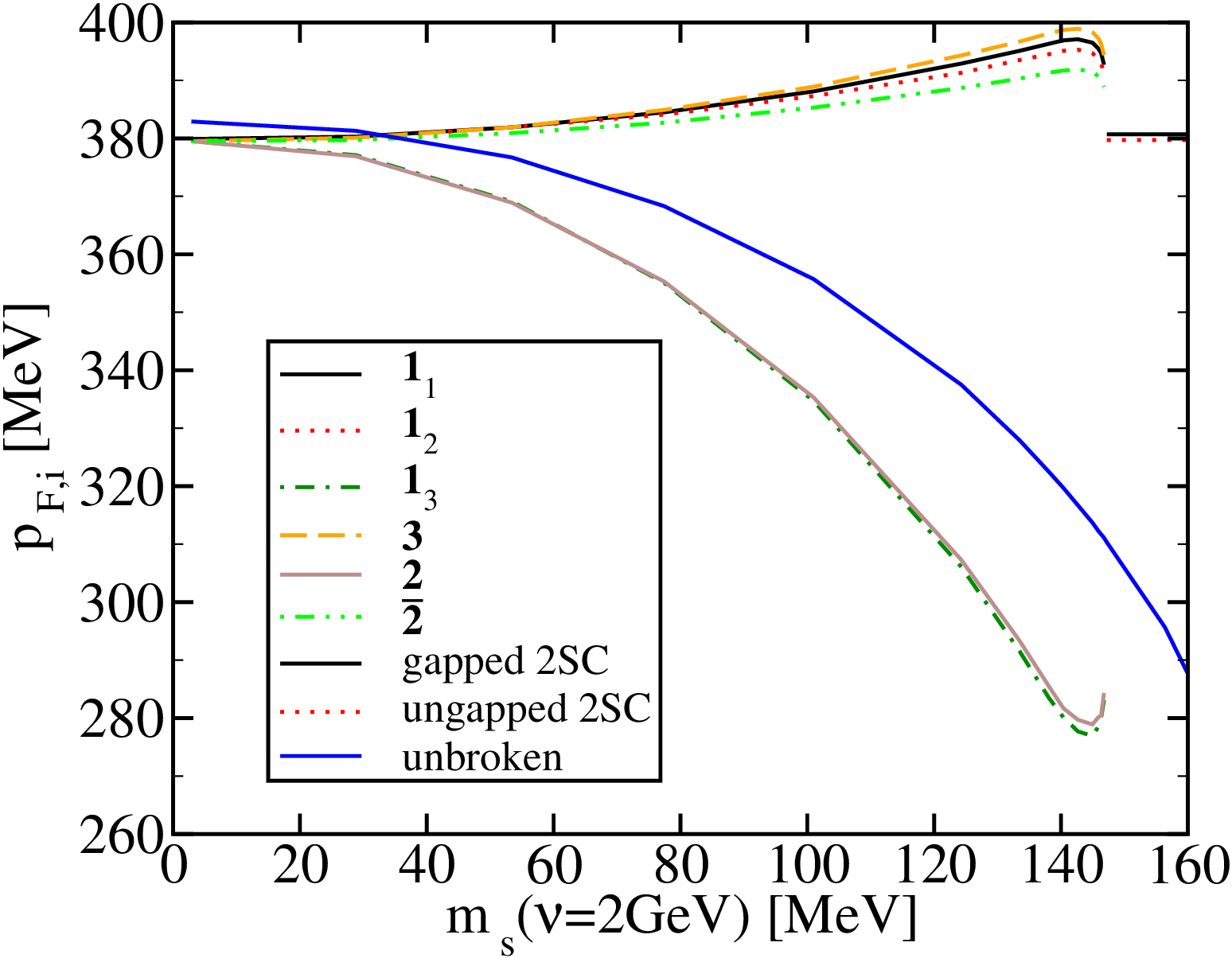}
  \hspace{.2cm}
  \includegraphics[width=8cm]{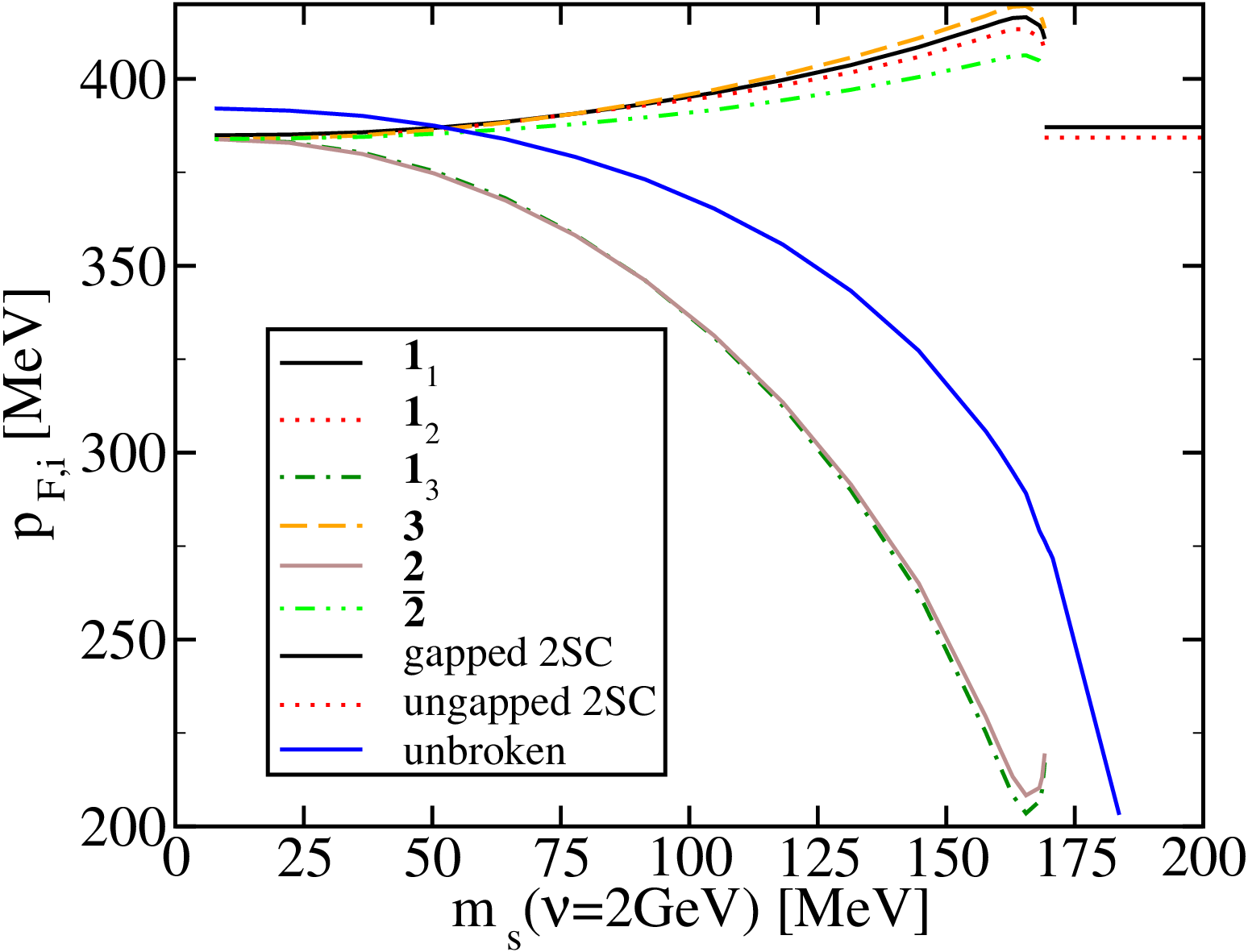}
  \caption{Fermi momenta in different channels (see text) at
  $\mu=400\,\mathrm{MeV}$ for the coupling $\alpha_{I}(k^{2})$ (left) and
  $\alpha_{II}(k^{2})$ (right).}
  \label{pFfigures}
\end{figure}

In Fig.\ \ref{pFfigures} the results for the Fermi momenta at a chemical
potential of $\mu=400\,\mathrm{MeV}$ as a function of the renormalized strange
quark current mass $m_{s}(\nu)$ at a renormalization scale $\nu=2\,\mathrm{GeV}$
are presented. For the CFL phase these are of course only plotted below the
critical value of the strange quark current mass. (For the definitions of the
different components see section~\ref{CFLdef}.) Above this critical value the
2SC phase is energetically preferred, and the three different Fermi momenta of
the 2SC phase are shown: For the gapped red and green, up- and down-quarks and
the ungapped blue up- and down-quarks, which are both independent of the
strange quark mass and for the decoupled strange quarks in the unbroken
phase (which is also displayed below the critical of $m_s$).

In the CFL phase the Fermi momenta are not monotonous functions of $m_s$. This
is due to the behavior of the vector self-energies and not visible in the
mass functions, {\it cf.\/} Fig.\ \ref{Msfigures}.

\subsection{Mass functions}
\label{massfuncts}
\begin{figure}[h!]
  \hspace{-.2cm}
  \includegraphics[width=8cm]{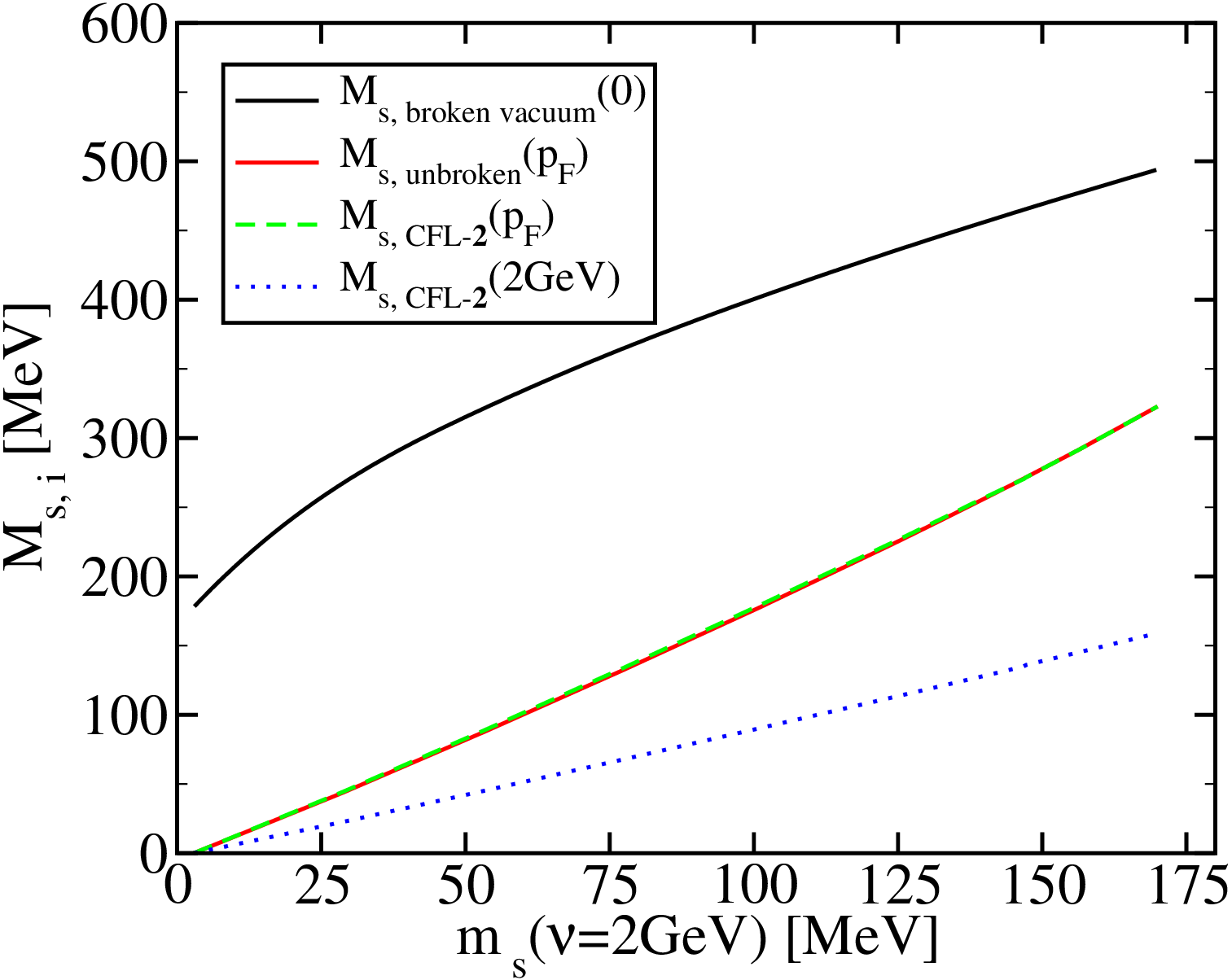}
  \hspace{.2cm}
  \includegraphics[width=8cm]{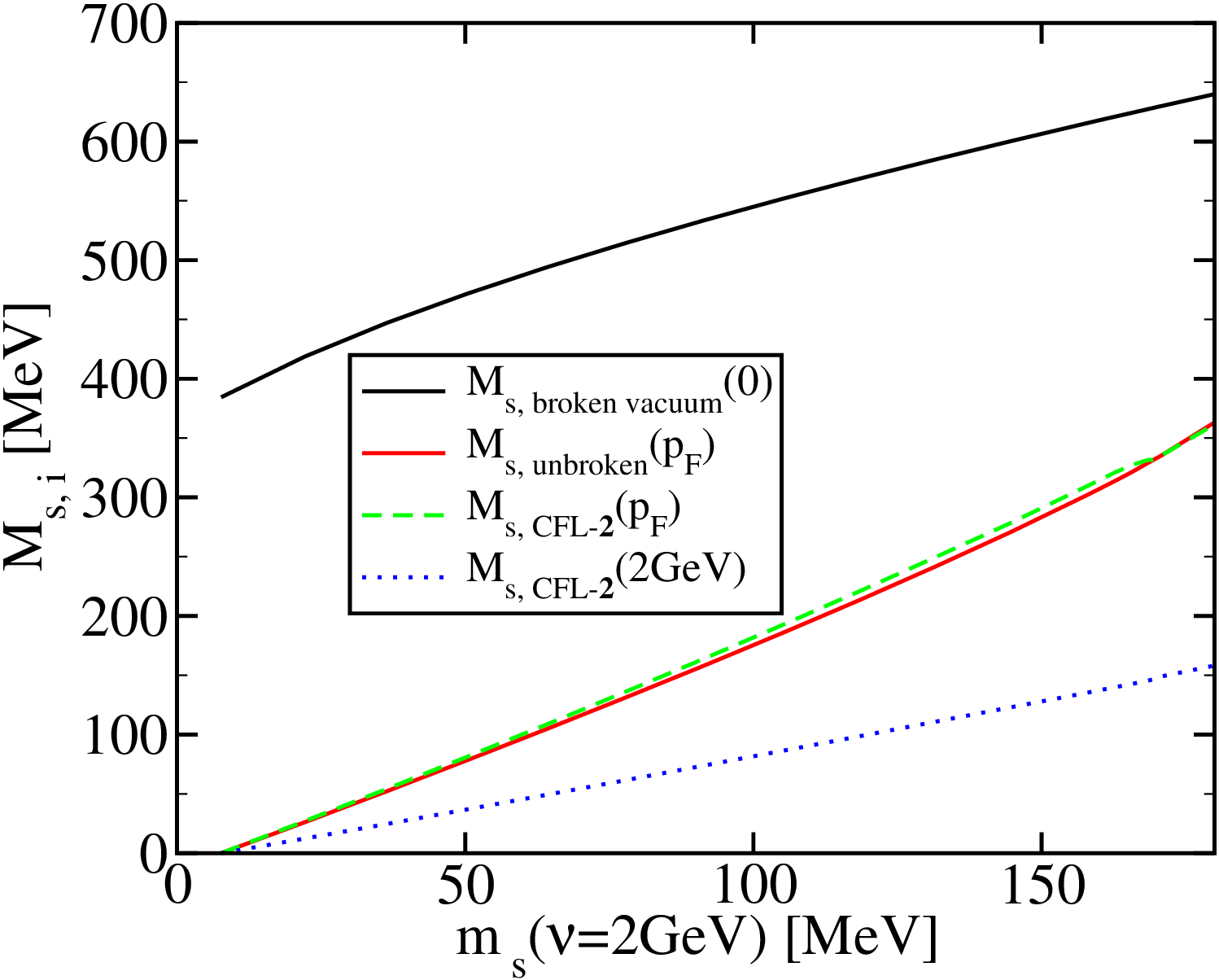}
  \caption{Mass functions at different values of three-momentum 
  (see text) as function of the
  renormalized strange quark mass at a chemical potential of
  $\mu=400\,\mathrm{MeV}$ for the coupling $\alpha_{I}(k^{2})$ (left) and
  $\alpha_{II}(k^{2})$ (right).}
  \label{Msfigures}
\end{figure}
As a function of $m_s$, we display in Fig.\ \ref{Msfigures} the results for
the constituent quark mass function in the vacuum at vanishing momenta
$M_{s,broken\,vacuum}(0)$, as well as the quark mass function in the unbroken
phase $M_{s,unbroken}(p_{F})$ and the doublet channel in the CFL phase
$M_{s,CFL-\pmb{2}}(p_{F})$, both at their respective Fermi momenta and for a
chemical potential of $\mu=400\,\mathrm{MeV}$. Furthermore, the quark mass
function in the unbroken phase $M_{s,unbroken}(2\,\mathrm{GeV})$ for the same
chemical potential at the renormalization scale $\nu=2\,\mathrm{GeV}$ is given.

Although the constituent quark mass in the vacuum $M_{s,broken\,vacuum}(0)$ is
very sensitive to the choice of the coupling, especially for small renormalized
strange quark masses, the mass functions at finite chemical potentials are not.
As explained in the previous section, this is due to the medium modification of
the coupling, which also leads to significantly smaller mass values at the
Fermi surface. Note furthermore that the values of the mass functions at the
Fermi surface in the chirally broken phase and the CFL phase are very close to
each other. This leads to the conclusion  that the dynamics near the Fermi
surface, where gapped and ungapped propagators strongly differ, are not
directly relevant for the dynamical mass generation. As expected, the mass
functions at the renormalization scale are comparable to the values in the
chirally broken vacuum, which confirms that $\nu$ is already sufficiently
above the scale of dynamical mass generation and the Fermi energy.

\subsection{Dependence of the gap functions on the renormalized strange-quark
current mass}
\label{gapmass}
\begin{figure}[h!]
  \hspace{-.2cm}
  \includegraphics[width=8cm]{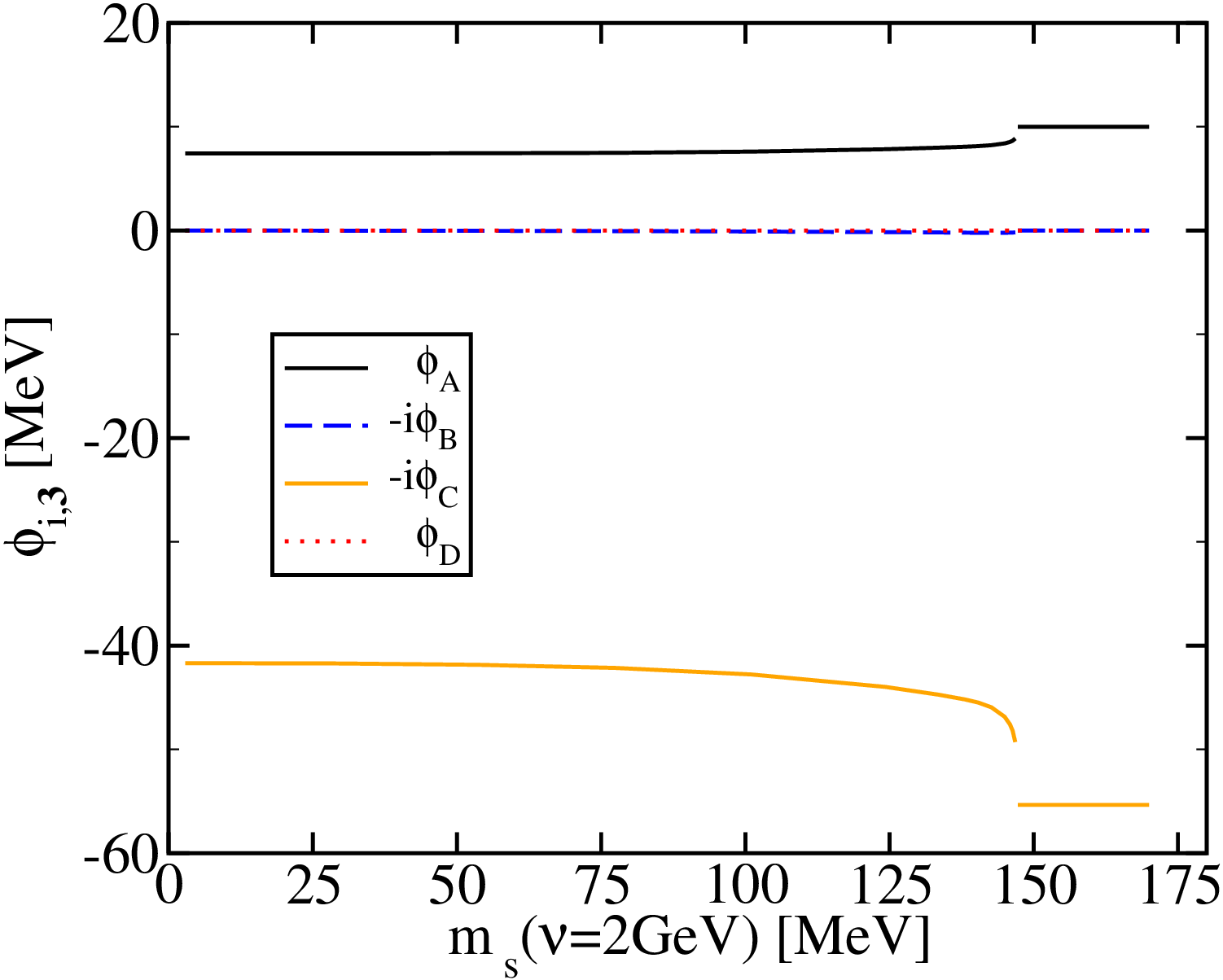}
  \vspace{.3cm}
  \hspace{.2cm}
  \includegraphics[width=8cm]{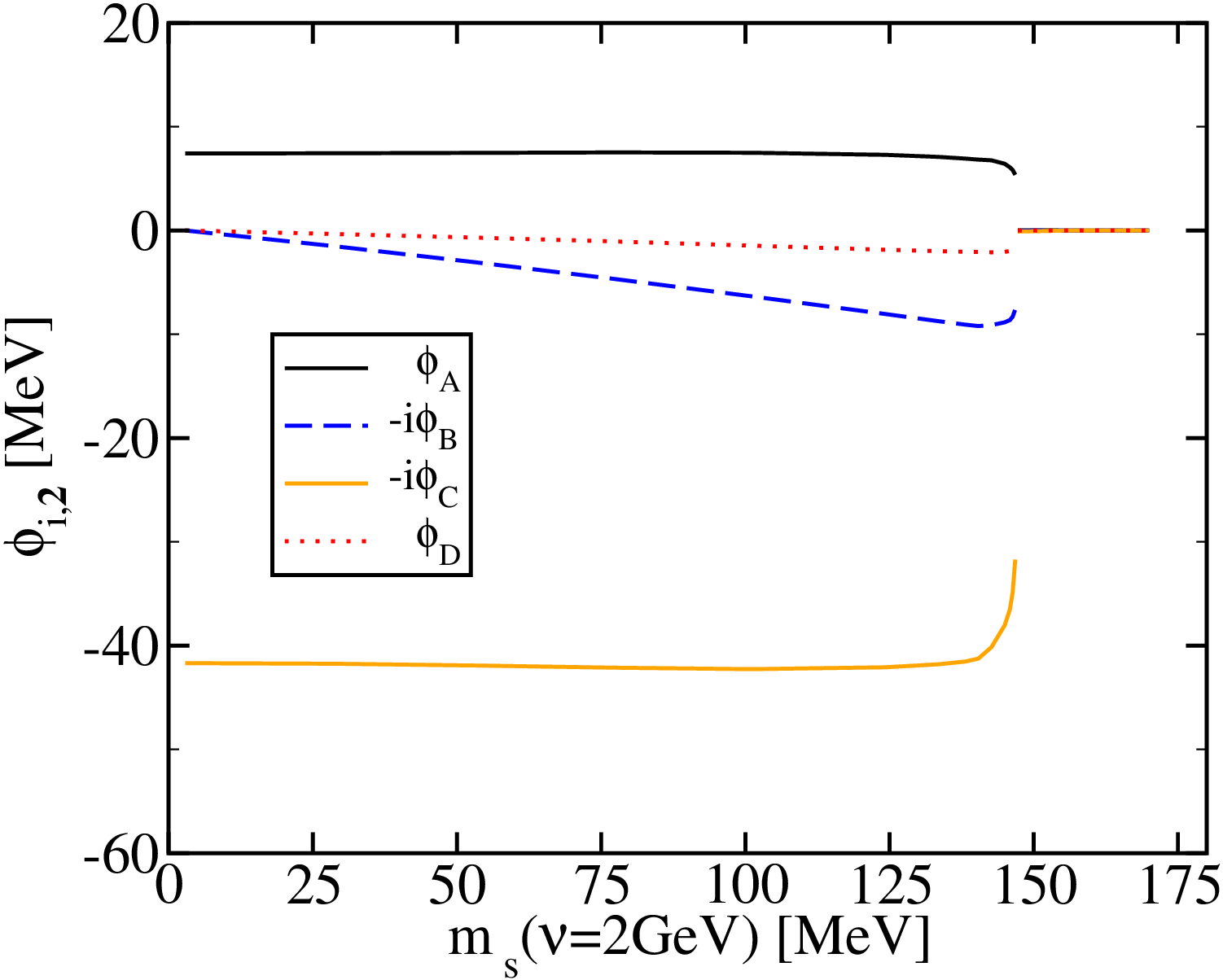}
  \hspace{-.2cm}
  \vspace{.3cm}
  \includegraphics[width=8cm]{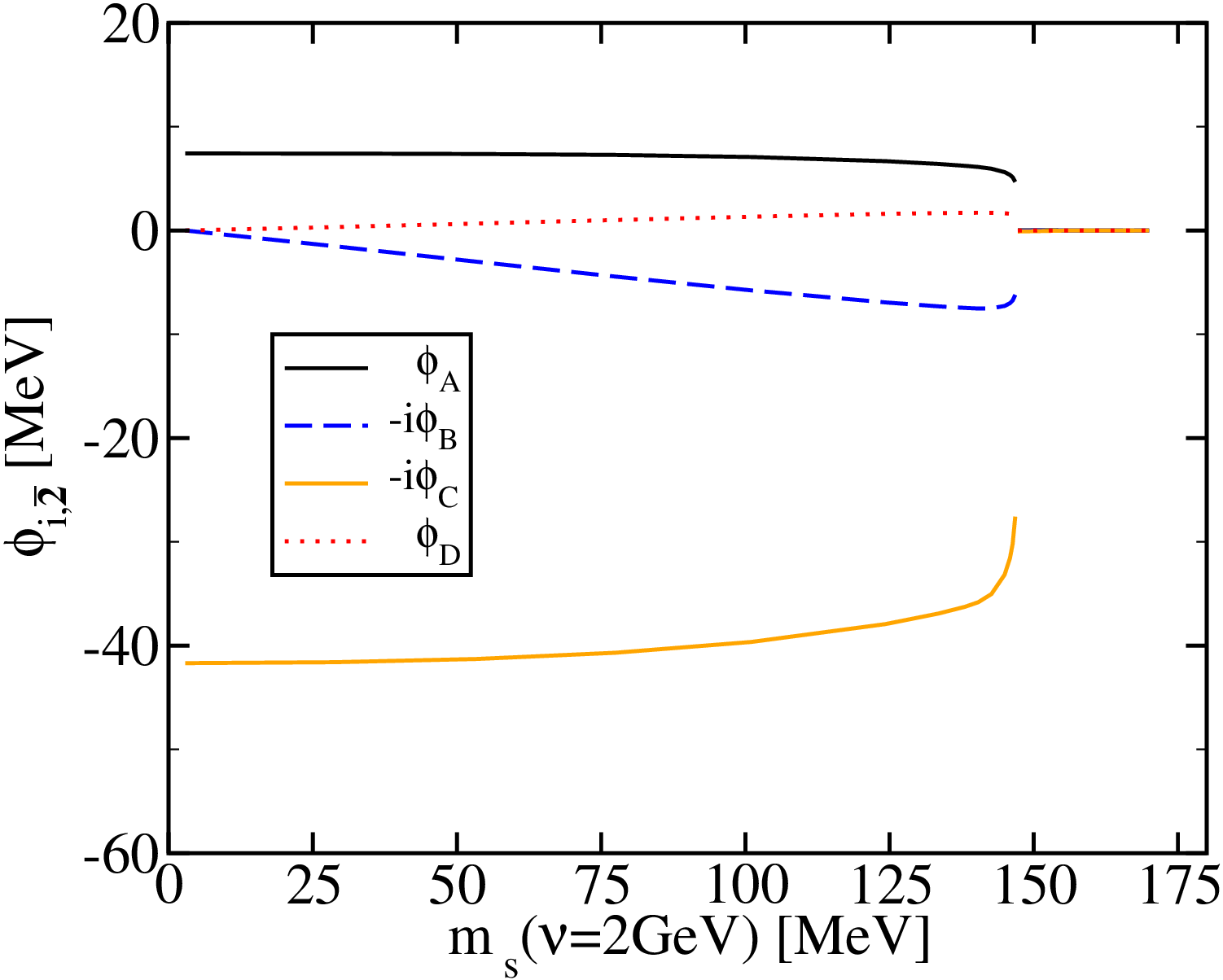}
  \hspace{.2cm}
  \vspace{.3cm}
  \includegraphics[width=8cm]{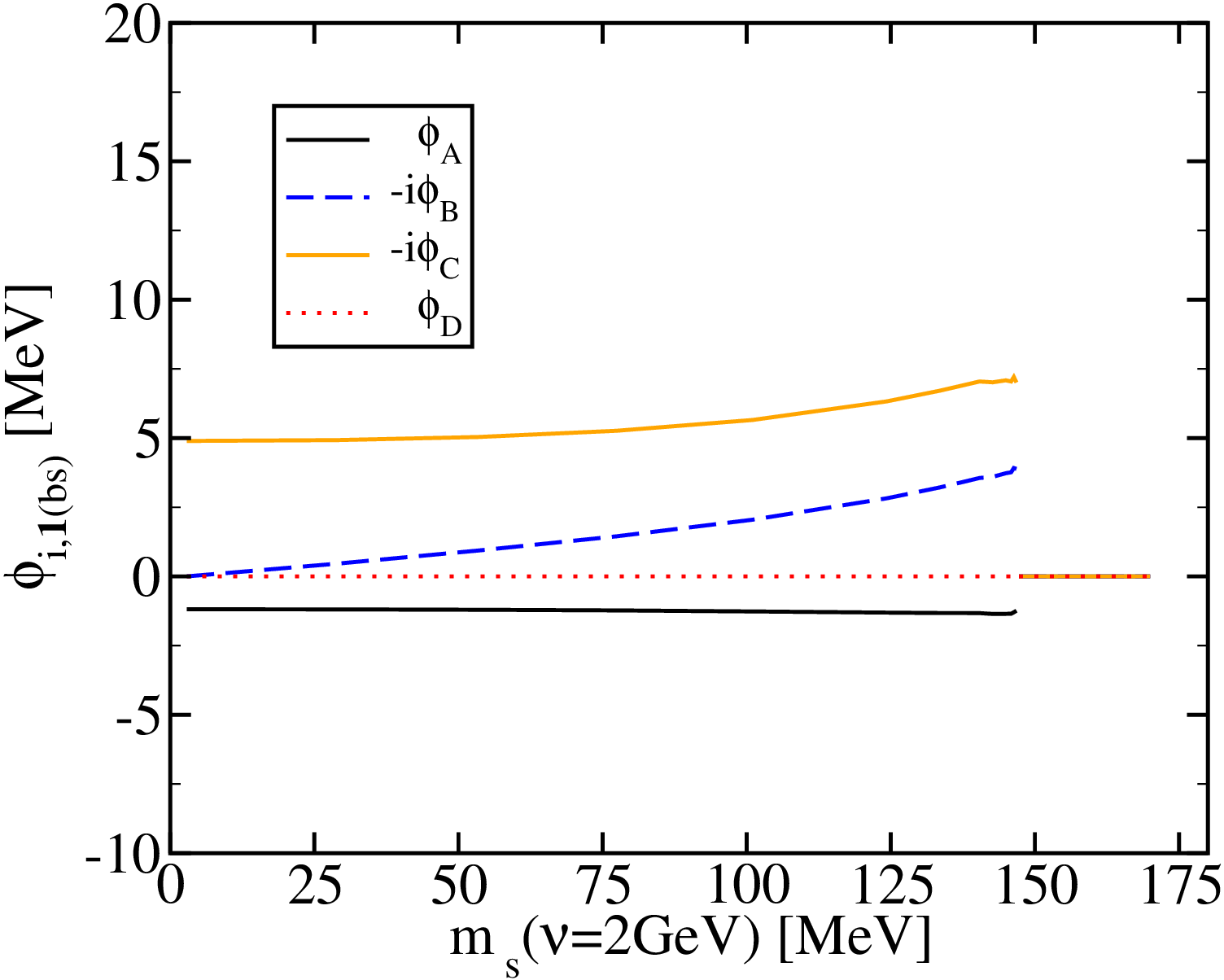}
  \hspace{-.2cm}
  \includegraphics[width=8cm]{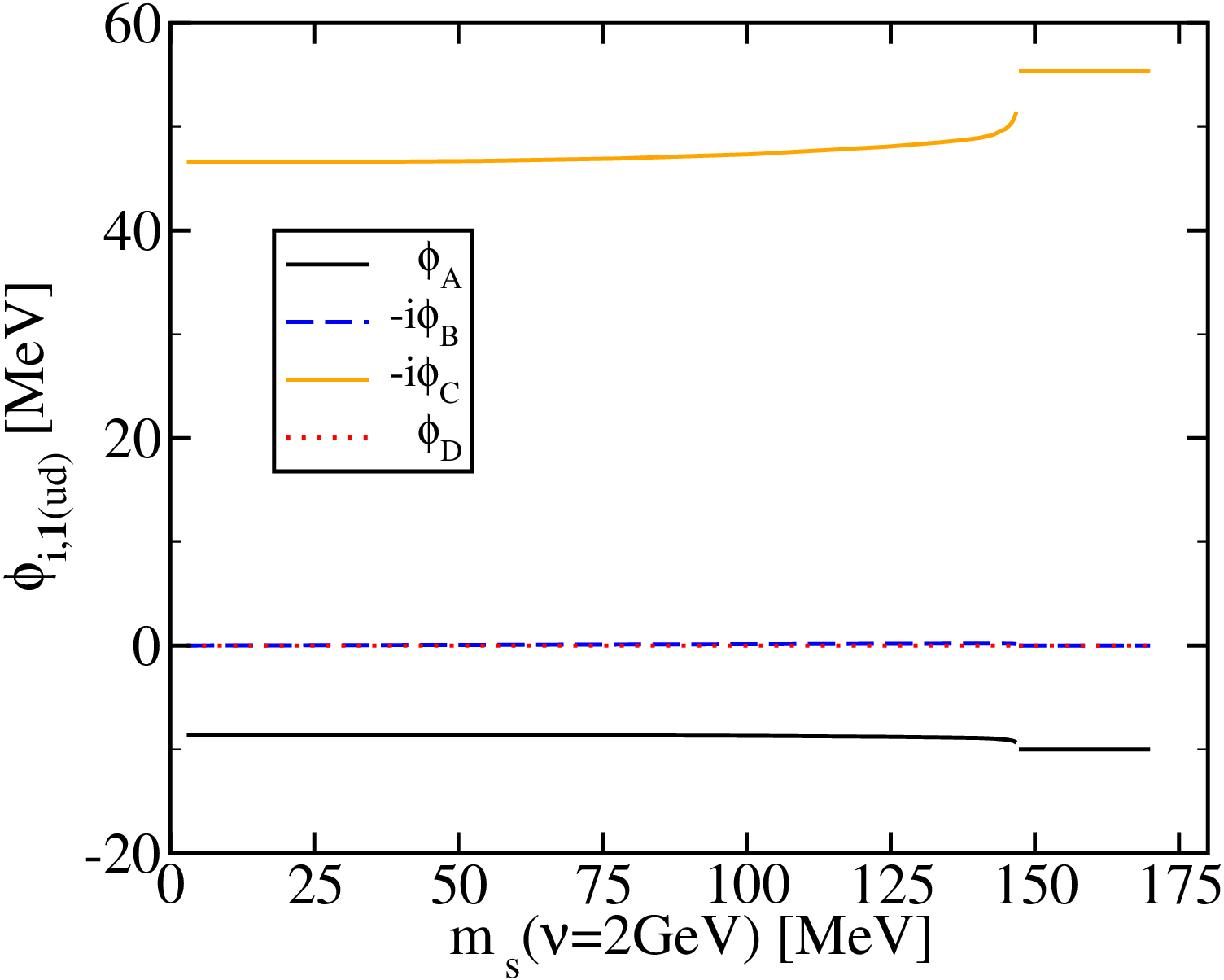}
  \hspace{.2cm}
  \includegraphics[width=8cm]{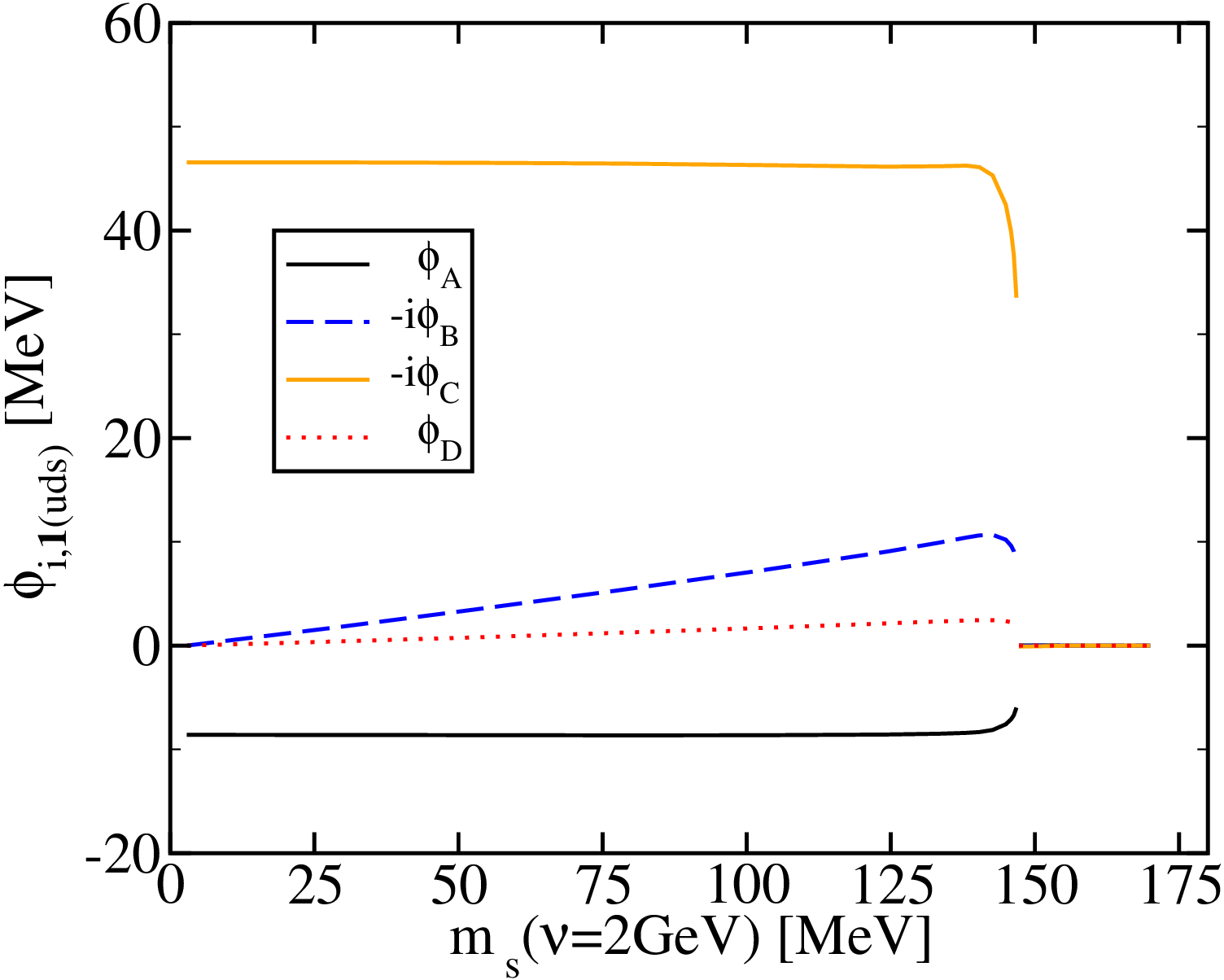}
  \caption{Gap functions at selected values of the three-momentum
  for different channels (see
  text) as a function of the renormalized strange-quark mass at a chemical
  potential of $\mu=400\,\mathrm{MeV}$ and for the coupling
  $\alpha_{I}(k^{2})$.}
  \label{phimsDSEfigs}
\end{figure}

\begin{figure}[h!]
  \hspace{-.2cm}
  \includegraphics[width=8cm]{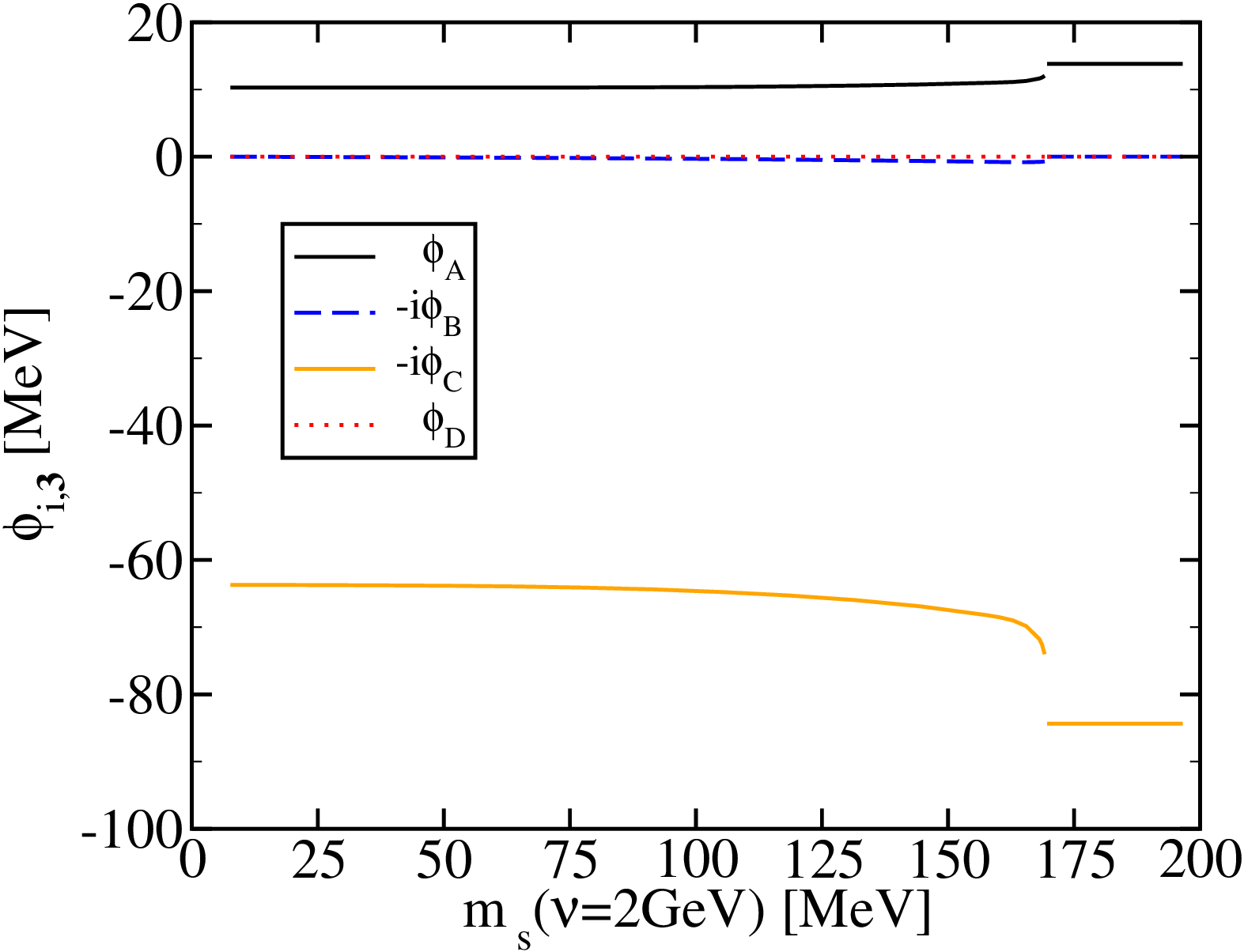}
  \vspace{.3cm}
  \hspace{.2cm}
  \includegraphics[width=8cm]{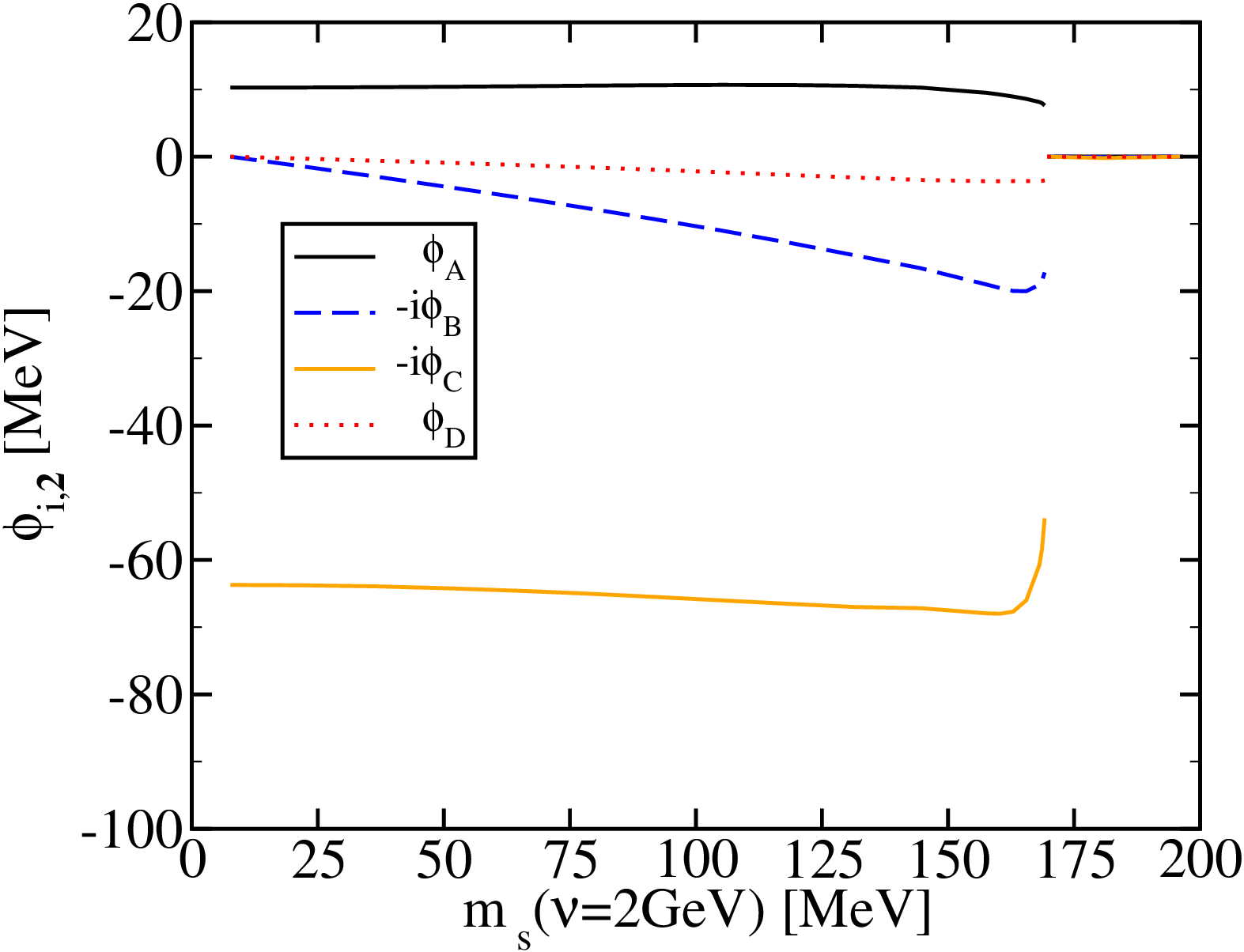}
  \vspace{.3cm}
  \hspace{-.2cm}
  \includegraphics[width=8cm]{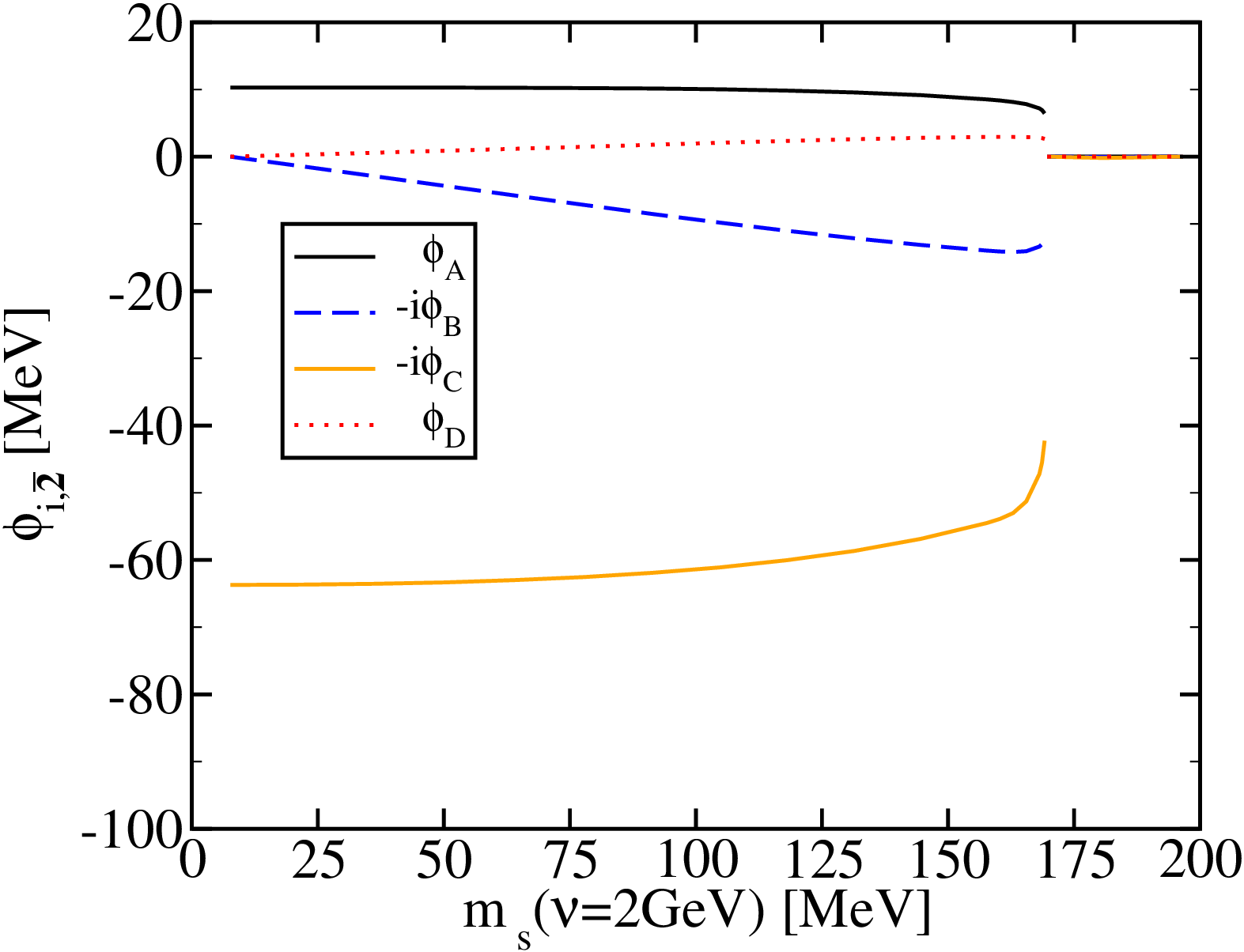}
  \vspace{.3cm}
  \hspace{.2cm}
  \includegraphics[width=8cm]{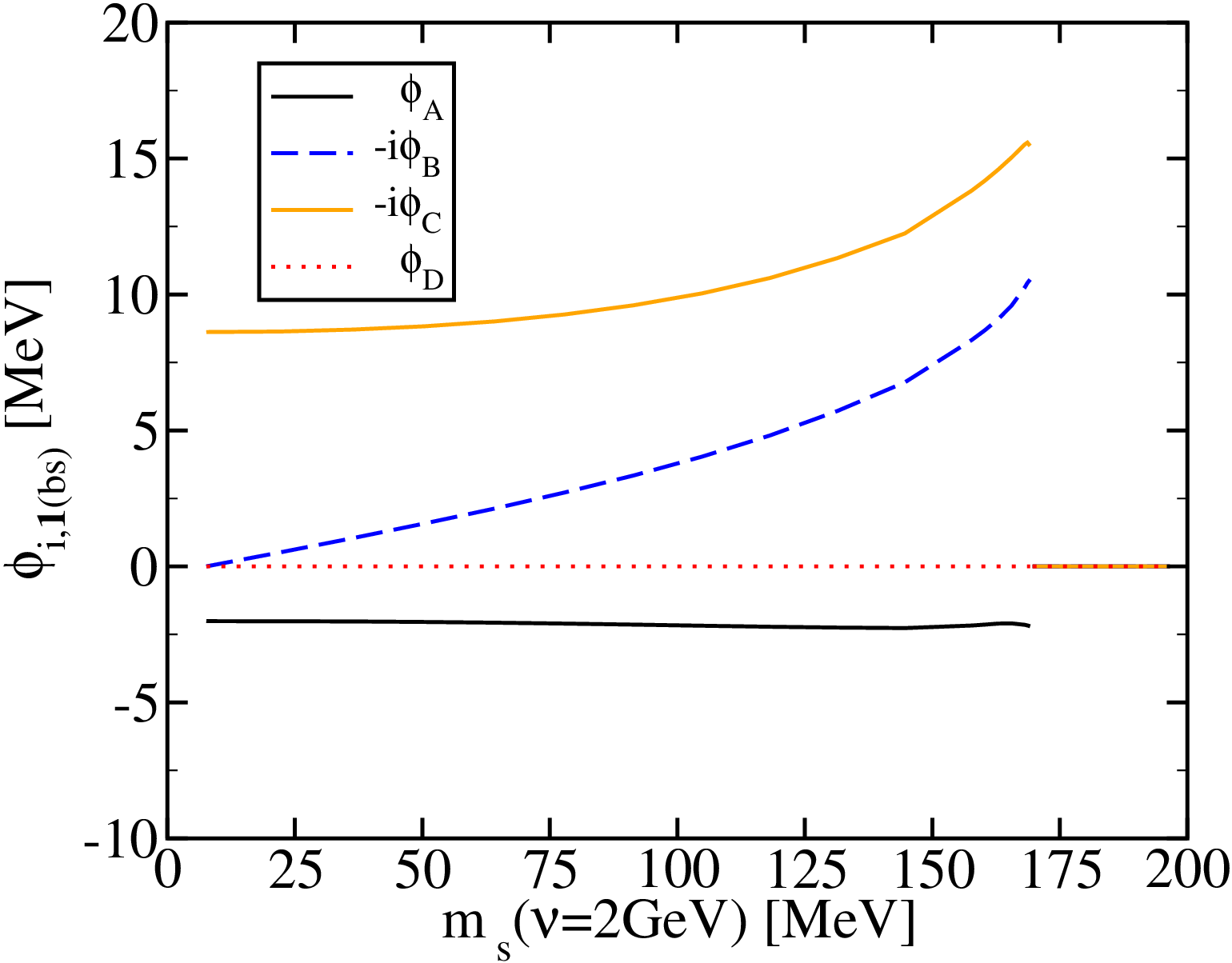}
  \hspace{-.2cm}
  \includegraphics[width=8cm]{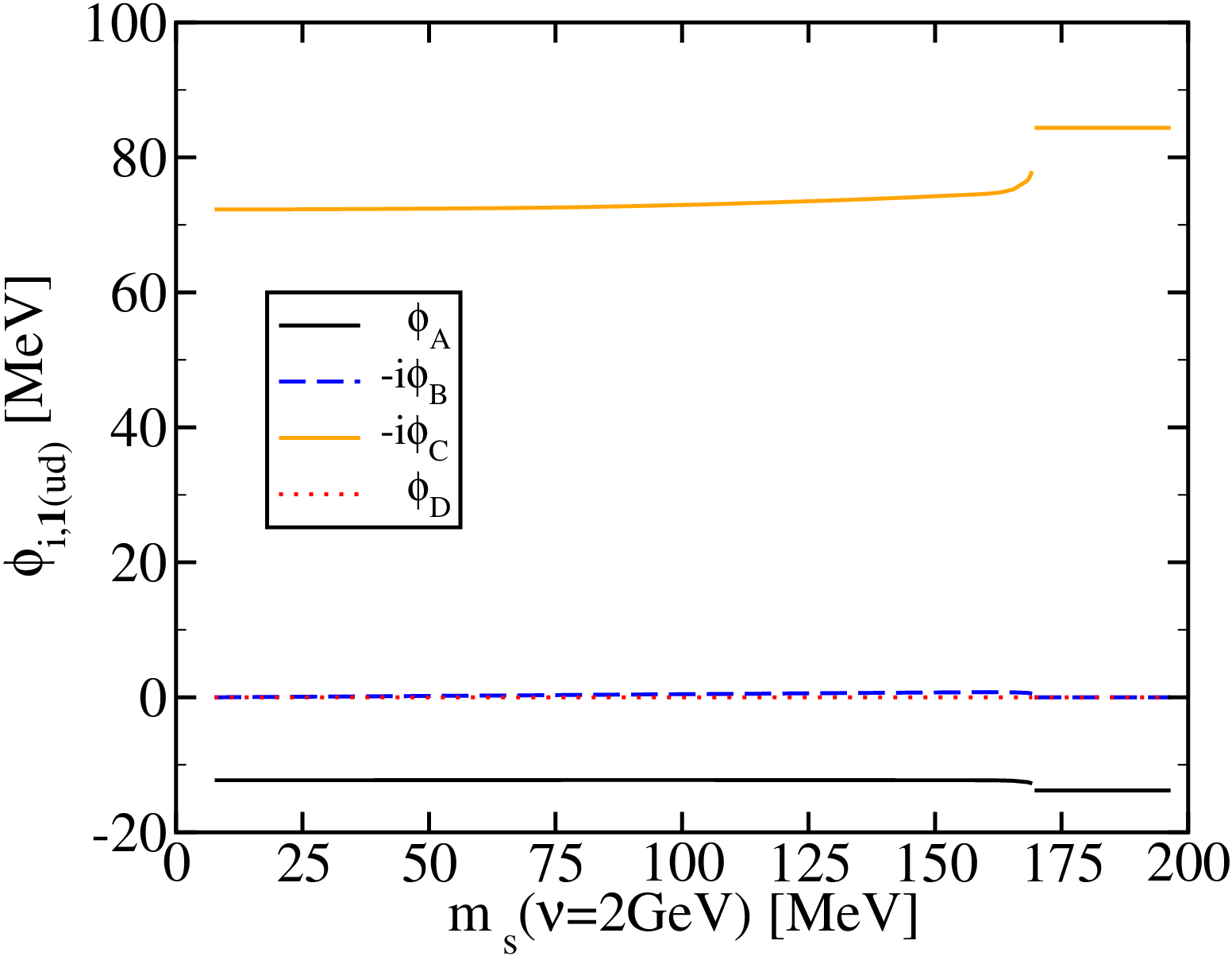}
  \hspace{.2cm}
  \includegraphics[width=8cm]{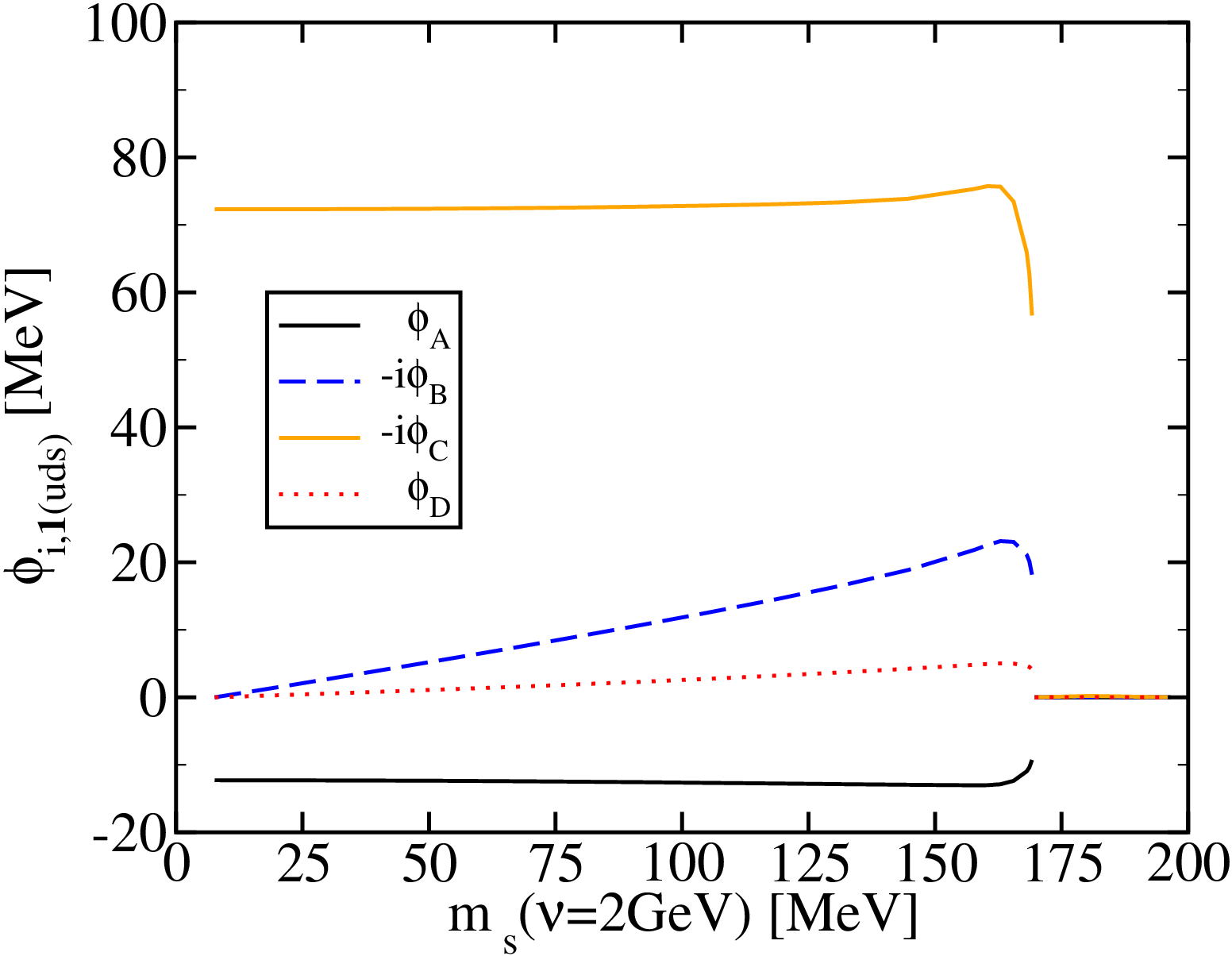}
  \caption{Gap functions at selected values of the three-momentum
  for different channels (see
  text) as function of the renormalized strange-quark mass at a chemical
  potential of $\mu=400\,\mathrm{MeV}$ and for the coupling $\alpha_{II}(k^{2})$.}
  \label{phimslatfigs}
\end{figure}

In this section we present results for the gap functions at the Fermi
surface, {\it i.e.\/} at $p_{4}=0$, and selected values of the three-momentum. 
The renormalization-point independent gap functions (\ref{phi}) for the
triplet, doublet and anti-doublet channel ({\it cf.\/} Eq.\ (\ref{TCFL})) are
evaluated at their corresponding Fermi momentum (see
section~\ref{CFLdef}). The functions $\phi_{i,\pmb{1}(ud)}$,
$\phi_{i,\pmb{1}(uds)}$ and $\phi_{i,\pmb{1}(bs)}$, corresponding to $M_{2}$,
$M_{4}$ and  $M_{3}$, given in the Appendix, are evaluated at
$(p_{F,\pmb{1}_{1}}+p_{F,\pmb{1}_{2}})/2$,
$(p_{F,\pmb{1}_{1}}+p_{F,\pmb{1}_{2}}+2p_{F,\pmb{1}_{3}})/4$ and
$p_{F,\pmb{1}_{3}}$, respectively. Due to this, the functions are considered
at momenta corresponding to the pairing quasiparticles. This also allows to
recover the results at the Fermi surface for the CFL phase in the chiral
limit. The corresponding numerical results are shown in Fig.\
\ref{phimsDSEfigs} for the coupling $\alpha_{I}(k^{2})$ and in Fig.\
\ref{phimslatfigs} for the coupling $\alpha_{II}(k^{2})$. Note again that due
to the phase choice the 
gap functions $\phi_{A,i}$ and $\phi_{D,i}$ are real and $\phi_{B,i}$ and
$\phi_{C,i}$ imaginary at $p_{4}=0$.

All functions, apart from $\phi_{i,\pmb{1}(bs)}$, evolve towards the
corresponding 2SC solution. However, $\phi_{i,\pmb{1}(bs)}$ already shows that
the transition must be first order. Furthermore, for the gap functions
$\phi_{i,\pmb{1}(ud)}$ and $\phi_{i,\pmb{3}}$, describing non-strange pairing,
$\phi_{B}$ and $\phi_{D}$ vanish, as expected. Finally, we see for the other
gap functions, relevant to strange pairing, that $\phi_{B}$ is the most
varying and $\phi_{D}$ is non-vanishing, which is in line with the
interpretation of the gap functions given in section~\ref{gapinter}. It again
is obvious that the sensitivity of the gap functions on the used coupling is
much weaker than those of the mass functions.

\subsection{Dependence of mass functions on the chemical potential}
\label{massfunctsmu}
\begin{figure}[h!]
  \hspace{-.2cm}
  \includegraphics[width=8cm]{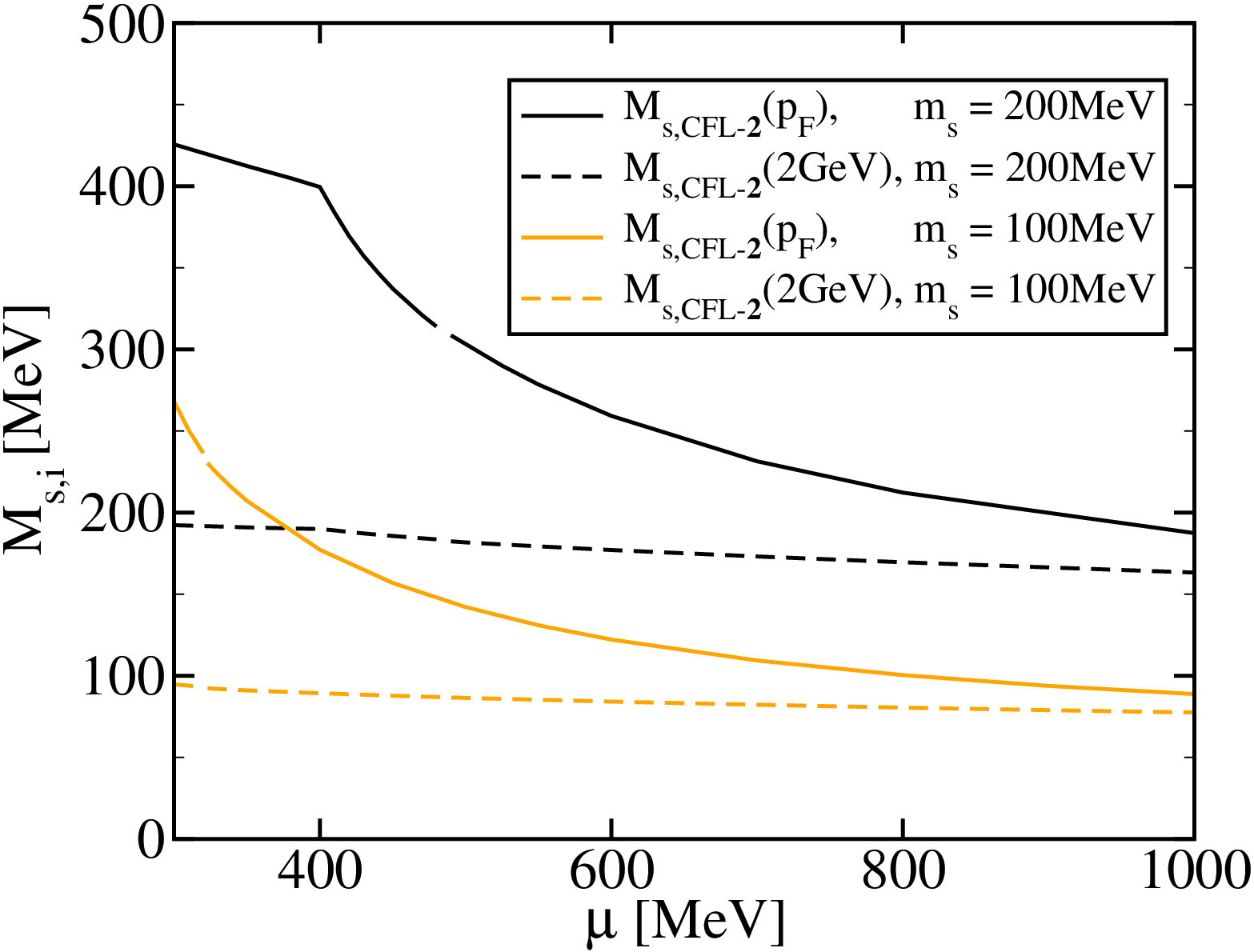}
  \hspace{.2cm}
  \includegraphics[width=8cm]{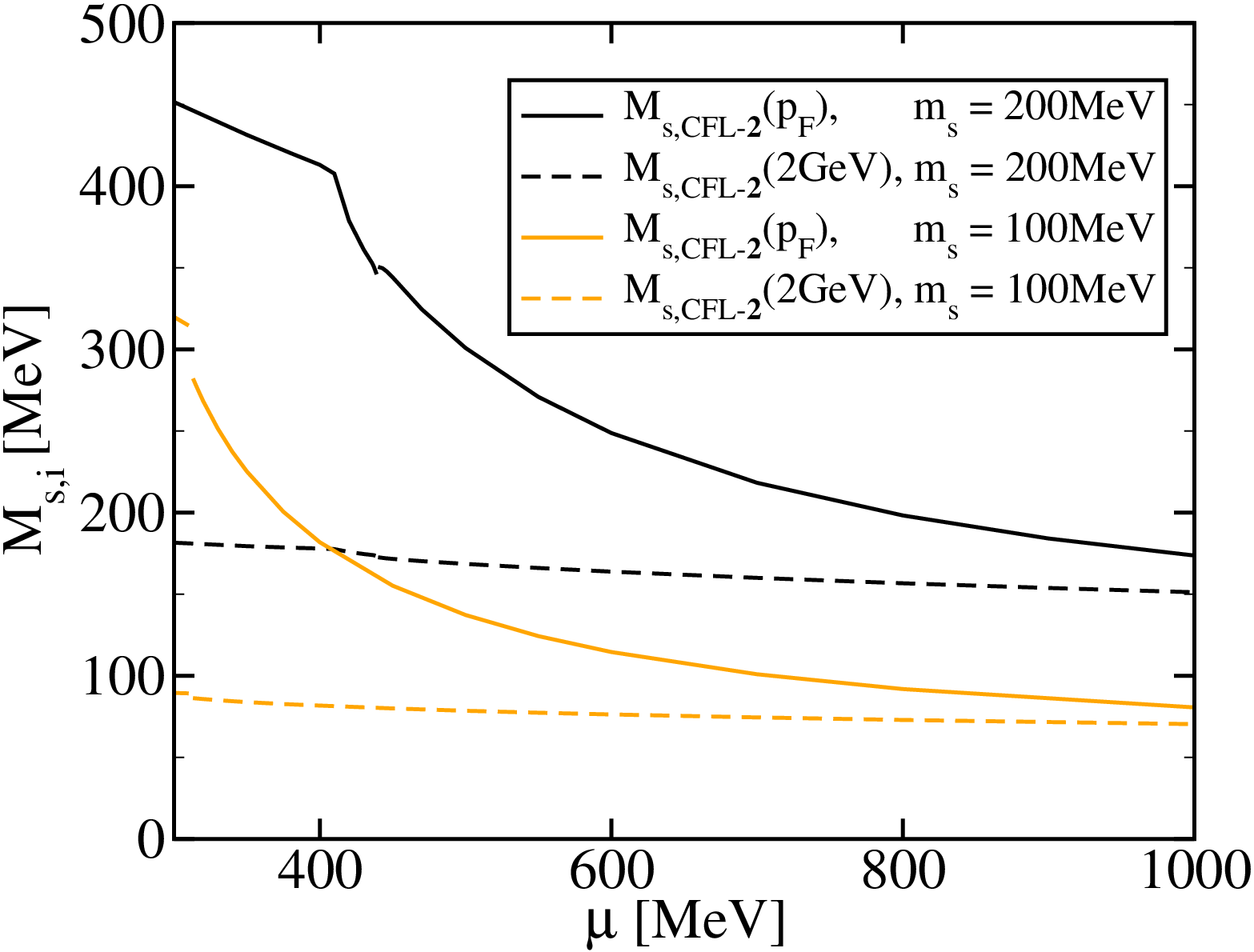}
  \caption{Mass functions on the Fermi surface and at a renormalization
  scale $\nu=2\,\mathrm{GeV}$ for fixed renormalized strange quark mass in the
  vacuum as function of the chemical potential for the coupling
  $\alpha_{I}(k^{2})$ (left) and $\alpha_{II}(k^{2})$ (right).}
  \label{massmufigs}
\end{figure}

We proceed by discussing the dependence on the chemical potential and  treat
the case of the mass functions first. Fig.\ \ref{massmufigs} shows these mass
functions in the doublet channel for two different renormalized strange-quark
current masses, evaluated at the Fermi momentum and at the renormalization
scale, respectively.

For small enough chemical potentials, the 2SC phase is preferred and the
doublet channel corresponds to decoupled strange quarks in the truncation used.
Depending on the value of the renormalized strange quark mass, the strange
quarks may not condense for small enough chemical potential,
{\it i.e.\/} not develop a Fermi surface. This can be seen from Fig.\
\ref{massmufigs} for a renormalized strange quark mass of
$m_{s}=200\,\mathrm{MeV}$. In this
case, we evaluate the mass function at vanishing momentum and the onset of
strange quark condensation is reflected as a kink in the curves.
With rising chemical potential, the system undergoes a phase transition into
the CFL phase. This effects the value of the mass function on the Fermi
surface only slightly and is shown as a gap in the plots. For the coupling
$\alpha_{II}(k^{2})$ at $m_{s}=100\,\mathrm{MeV}$ we find a direct transition
of non-condensed strange quarks into the CFL phase.

The values of the mass functions at the renormalization scale only show a
slight dependence on the chemical potential and are comparable to those in
the chirally broken vacuum, which again reflects the fact that the
renormalization  scale is well above the dynamical chiral symmetry breaking
scale and the chemical potential. On the other hand, the values of the mass
functions at the Fermi surface already at a chemical potential of $\mu
=1\,\mathrm{GeV}$ are close to their values at the renormalization scale. As a
result, dynamical chiral breaking is suppressed and the mass function is only
weakly dependent on the momentum below the renormalization scale.

\subsection{Dependence of gap functions on the chemical potential}
\label{gapmu}
\begin{figure}
  \hspace{-.2cm}
  \includegraphics[width=8cm]{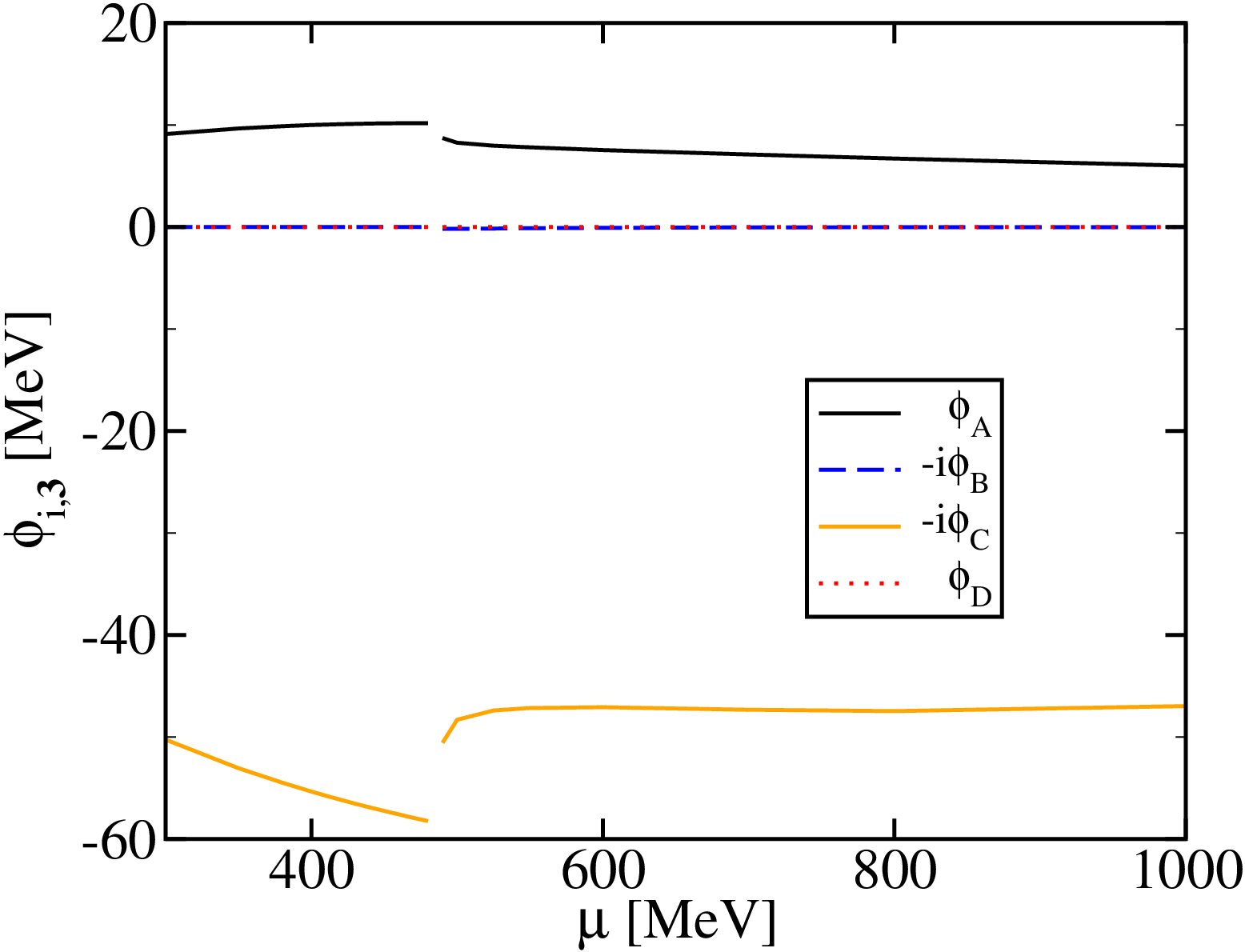}
  \vspace{.3cm}
  \hspace{.2cm}
  \includegraphics[width=8cm]{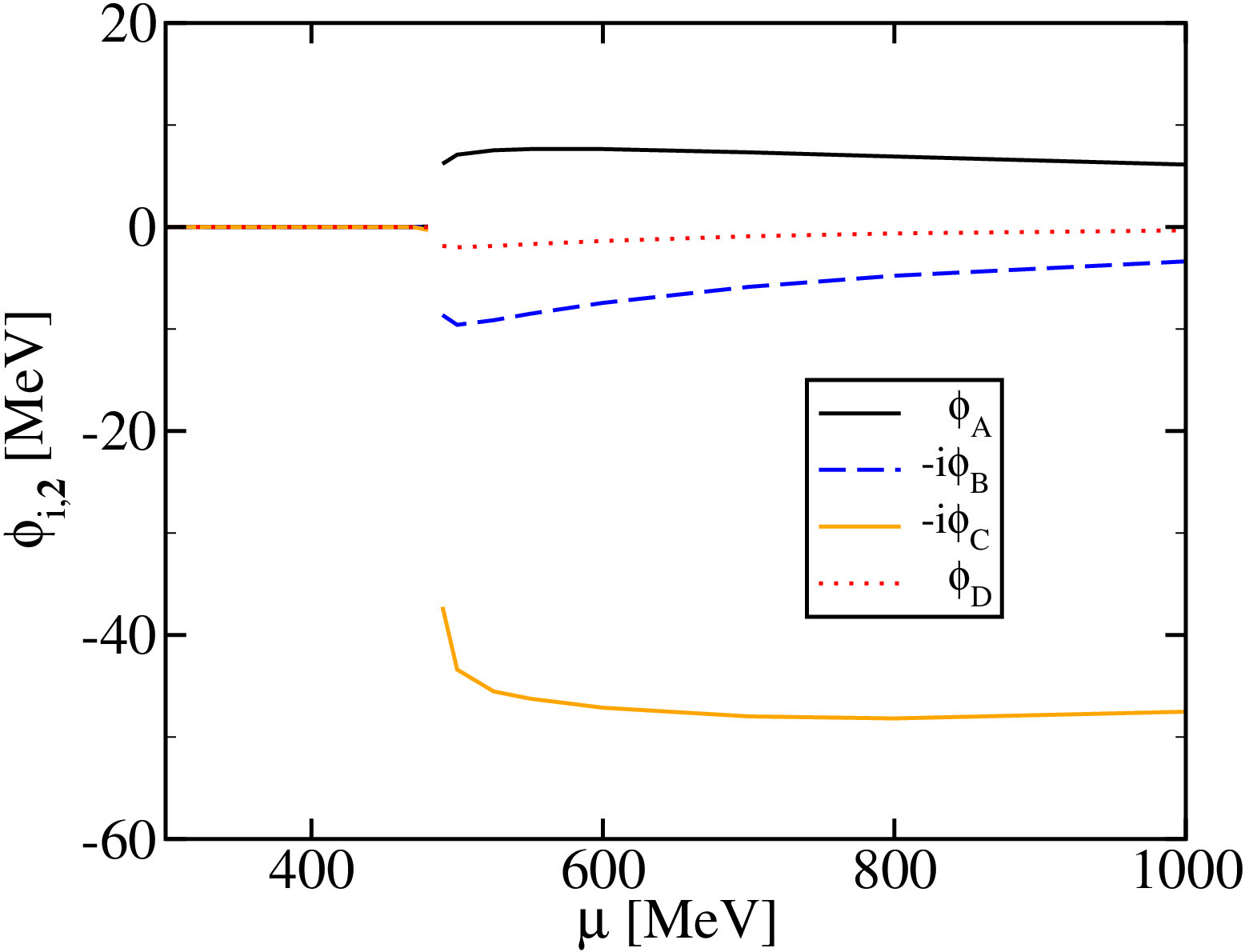}
  \vspace{.3cm}
  \hspace{-.2cm}
  \includegraphics[width=8cm]{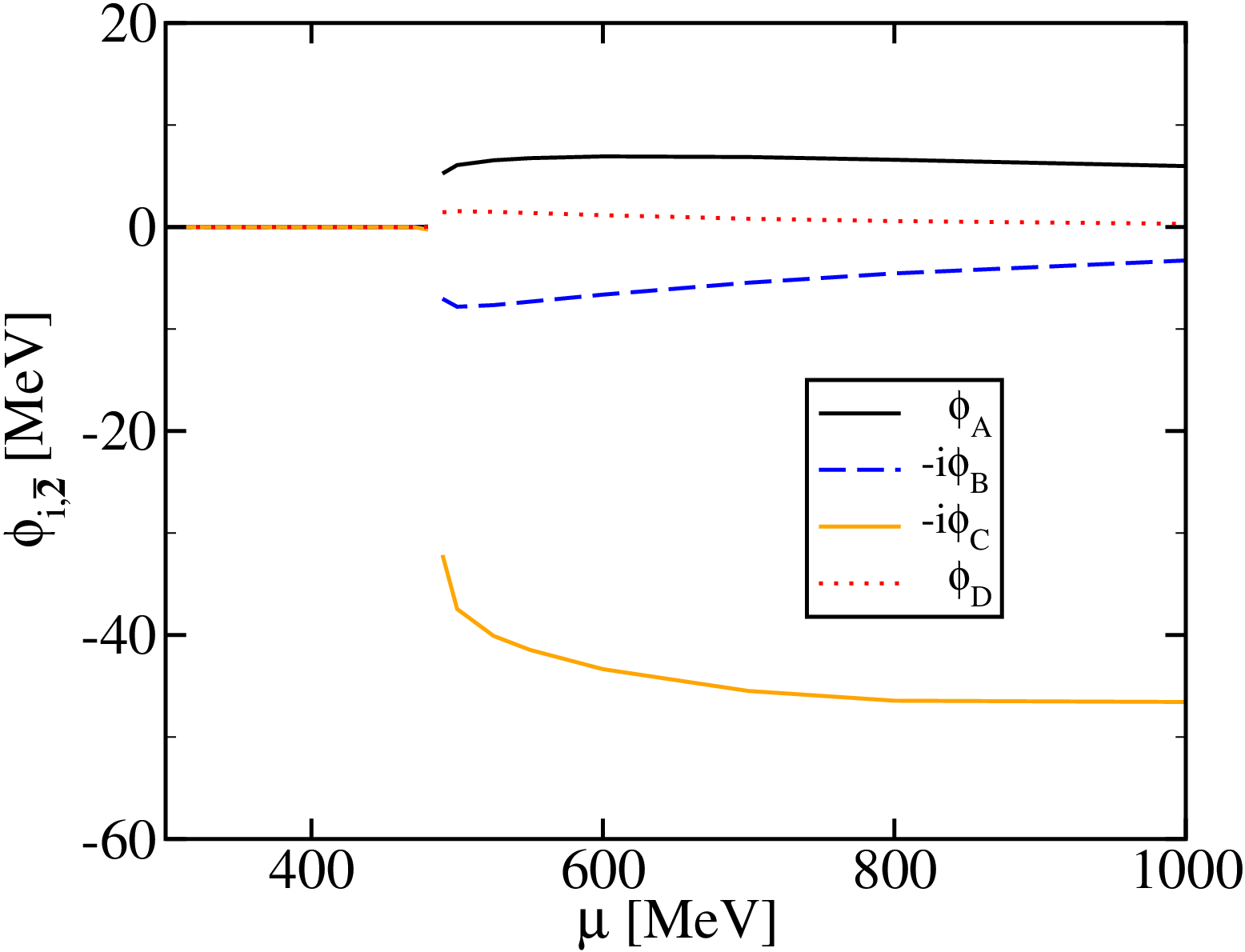}
  \vspace{.3cm}
  \hspace{.2cm}
  \includegraphics[width=8cm]{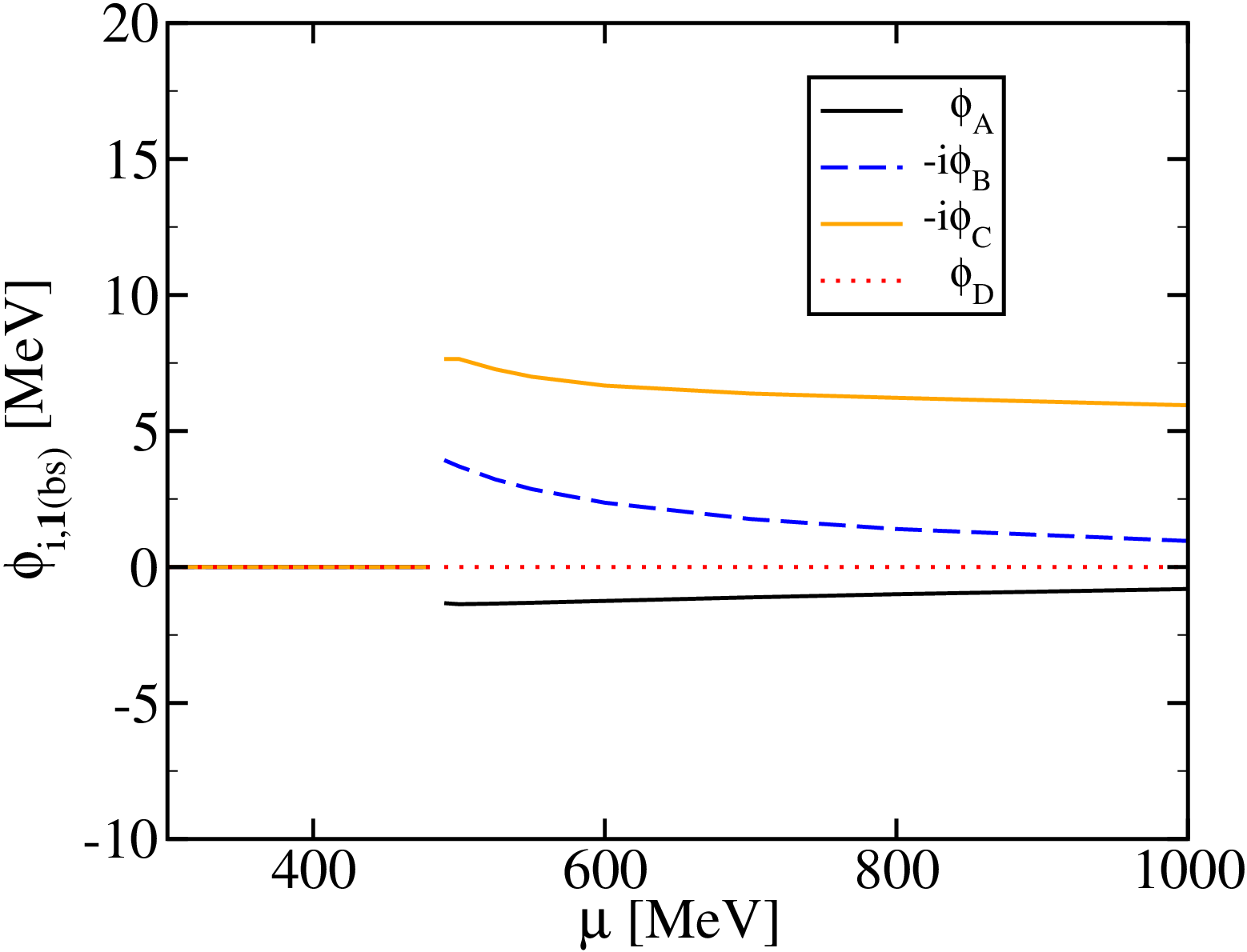}
  \hspace{-.2cm}
  \includegraphics[width=8cm]{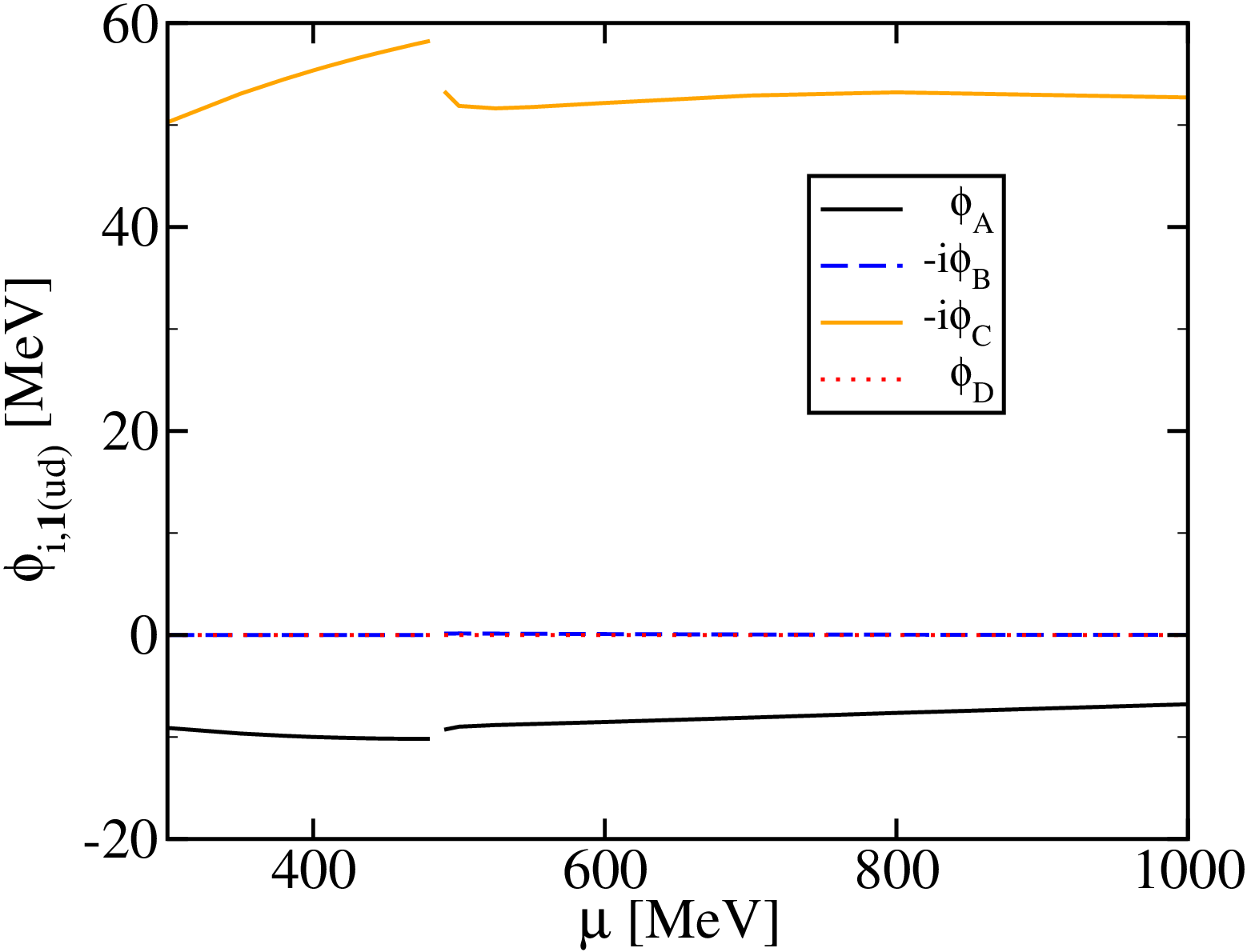}
  \hspace{.2cm}
  \includegraphics[width=8cm]{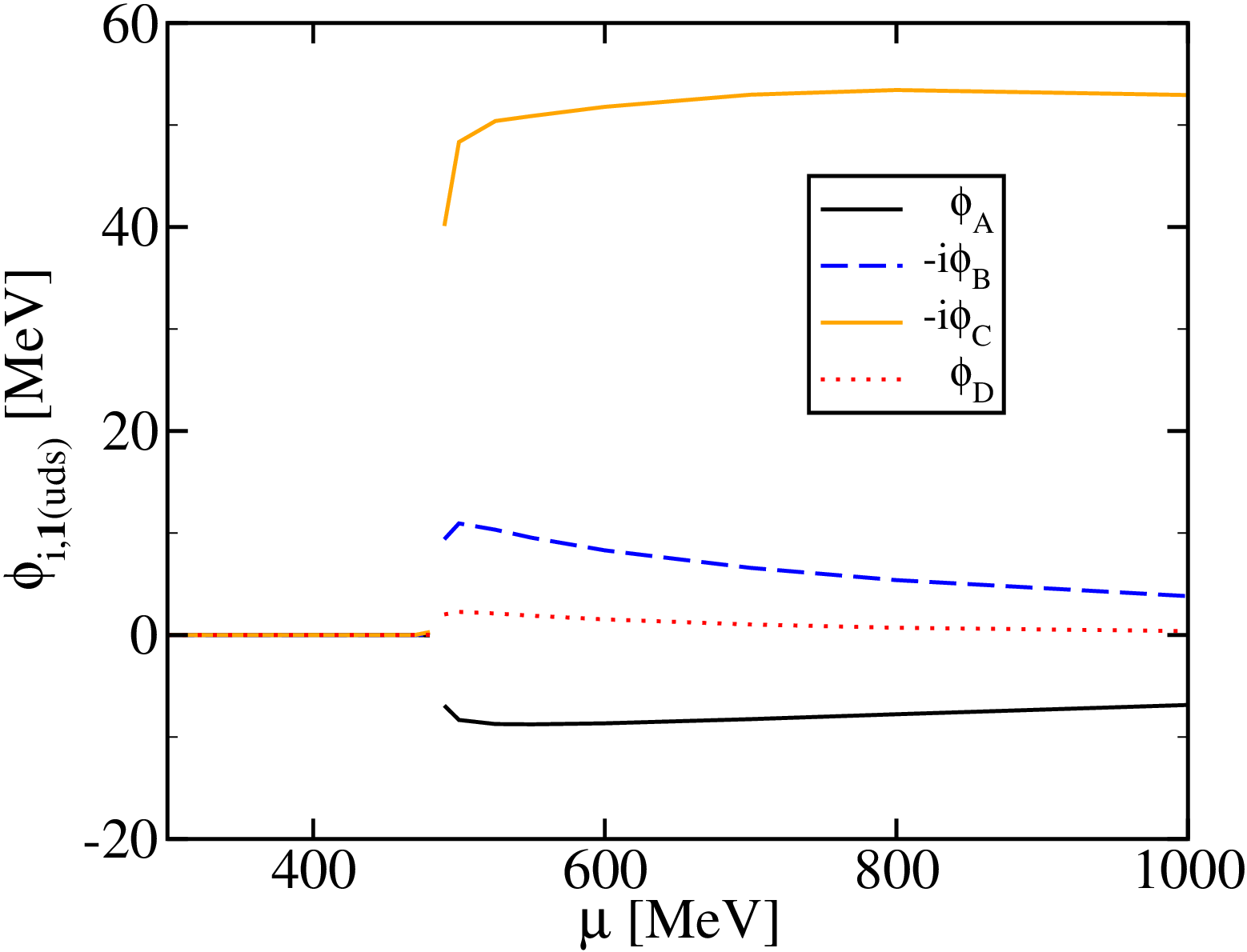}
  \caption{Gap functions on the Fermi surface for different channels (see
    text) at $m_{s}(\nu=2\,\mathrm{GeV})=200\,\mathrm{MeV}$ as a function of
    chemical potential for the coupling $\alpha_{I}(k^{2})$.}
  \label{phimufigs}
\end{figure}

For completeness we also present results for the dependence of the
gap functions on the chemical potential. In Fig.\ \ref{phimufigs} we show
the results for the gap functions in the CFL phase at a renormalized
strange quark mass of $m_{s}(\nu=2\,\mathrm{GeV})=200\,\mathrm{MeV}$ and for the
coupling $\alpha_{I}(k^{2})$. As described above, the 2SC phase is preferred
for smaller chemical potentials and we find again a visible jump in the
$\phi_{i,\pmb{1}(ud)}$ functions at a certain chemical potential. The gap
functions in the CFL phase are remarkably insensitive in the chemical
potential. Only the $\phi_{B}$ and $\phi_{D}$ functions evolve towards zero,
which again reflects that the relevant values of the mass functions also
become smaller.

\subsection{Pressure difference and critical strange-quark mass}
\label{mscrit}
\begin{figure}[h!]
  \hspace{-.2cm}
  \includegraphics[width=8cm]{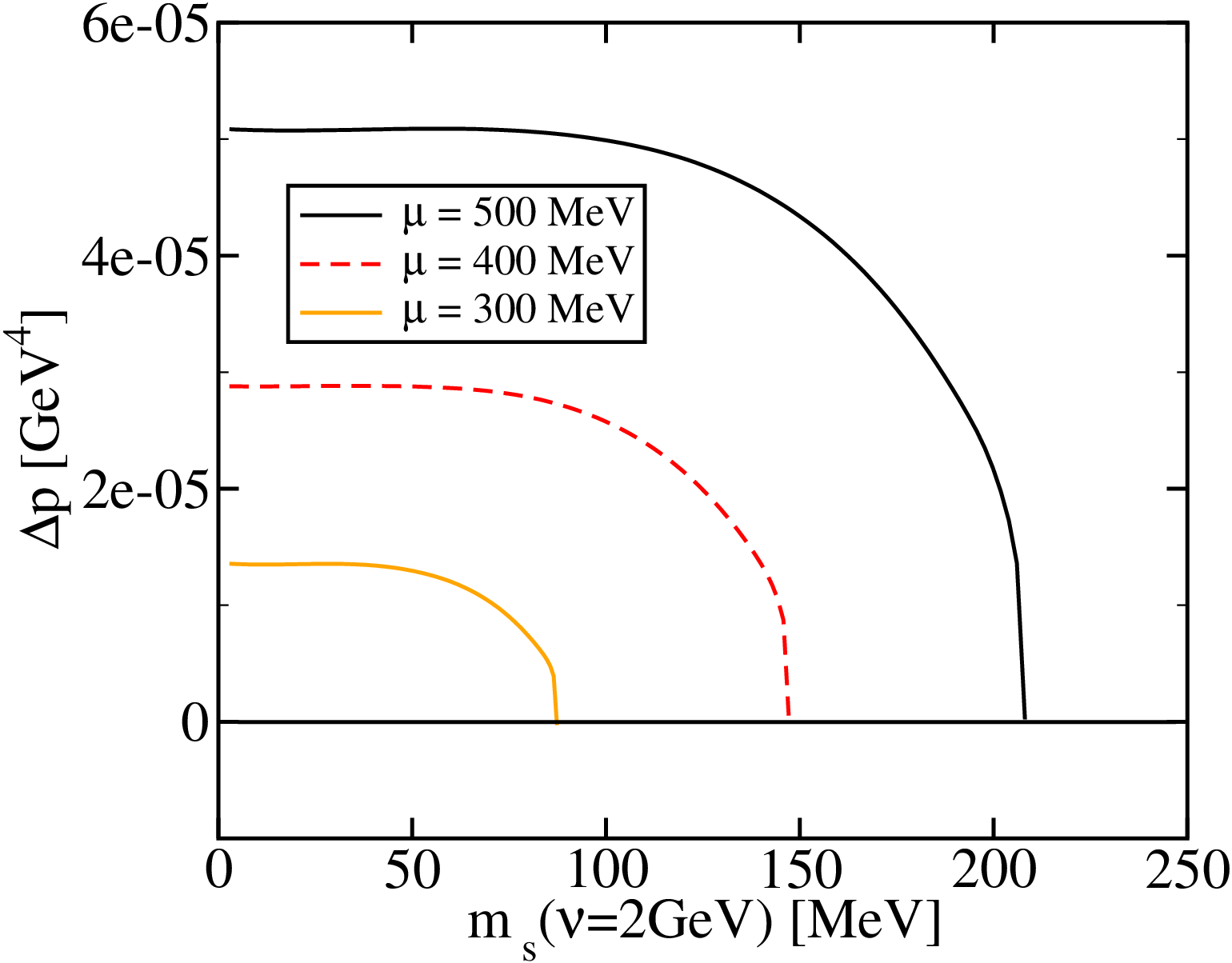}
  \hspace{.2cm}
  \includegraphics[width=8cm]{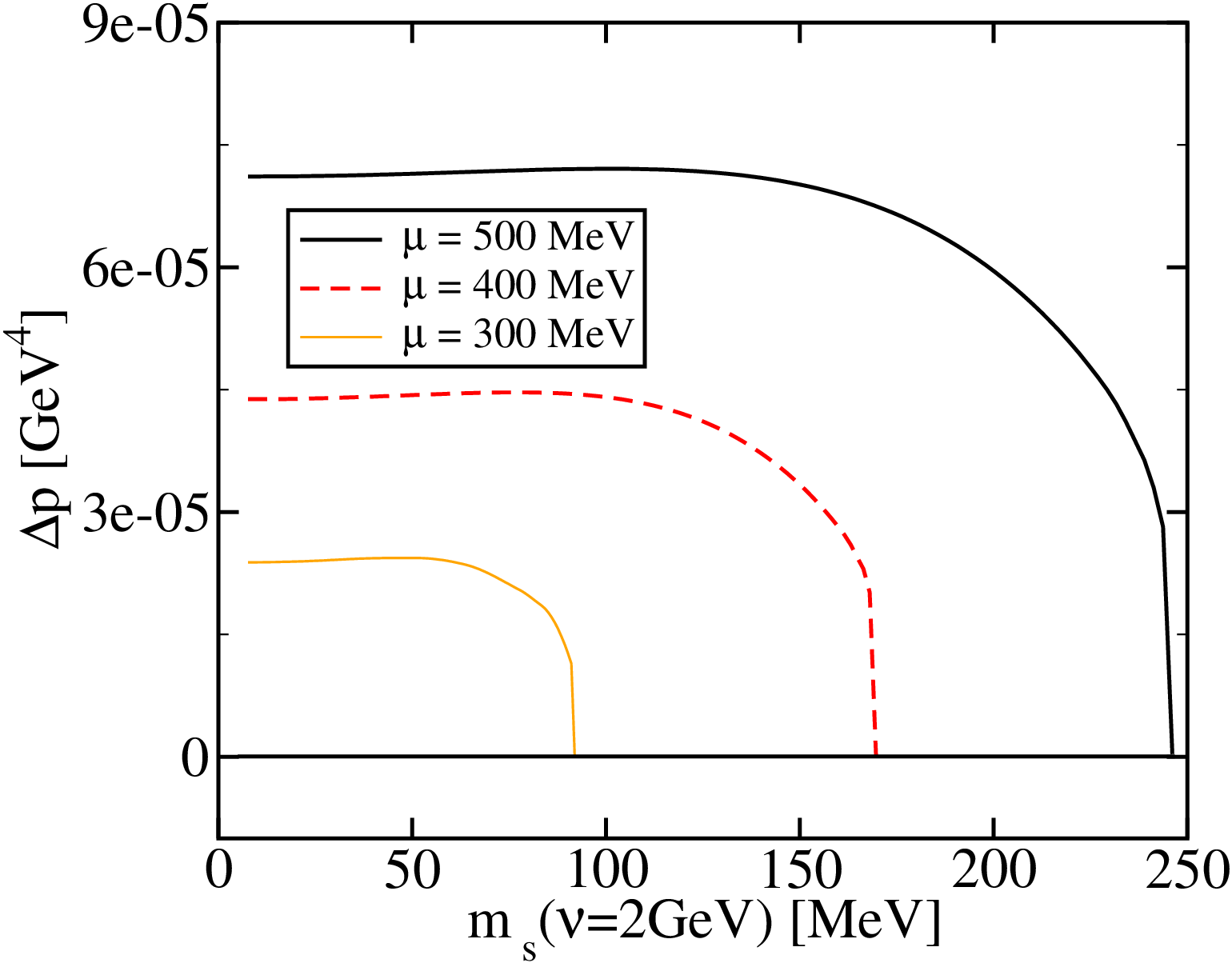}
  \caption{Pressure difference between 2SC and CFL phase as function of the
  renormalized strange quark mass at chemical potentials of
  $\mu=300\,\mathrm{MeV}, 400\,\mathrm{MeV}, 500\,\mathrm{MeV}$
  for the coupling $\alpha_{I}(k^{2})$ (left) and $\alpha_{II}(k^{2})$
  (right).}
  \label{Dpms}
\end{figure}

We now turn to the main result of this investigation: The determination of the
critical value of the strange-quark current mass. Above this mass the Fermi
surfaces are so far separated that pairing of up and down quarks with strange
quarks is no longer energetically preferred. Based on the CJT-formalism we
determine the pressure difference of CFL and 2SC using Eq.\
(\ref{CJTeqn}). The results as a function of the renormalized strange quark
current mass for different chemical potentials and for the couplings employed
is shown in Fig.\ \ref{Dpms}. One sees, as expected, that for small masses the
CFL phase is
preferred, and that there is a critical value of the strange-quark mass where
the CFL phase becomes energetically disfavored.

We like to emphasize that we are no longer able to find a solution for the CFL
phase in case the 2SC phase becomes favored. This is considered as a
consequence of the numerical method to solve the truncated DSE being the
functional derivative of the truncated CJT action. It turns out that we always
only find the global minimum of the CJT action as long as the local minimum is
not protected by a higher symmetry. The latter is the case for the 2SC phase,
if the CFL one is preferred. Nevertheless we can judge from the behavior of the
gap functions that the transition is first order.

\begin{figure}[h!]
  \begin{center}
    \hspace{-1.5cm}
    \includegraphics[width=10cm]{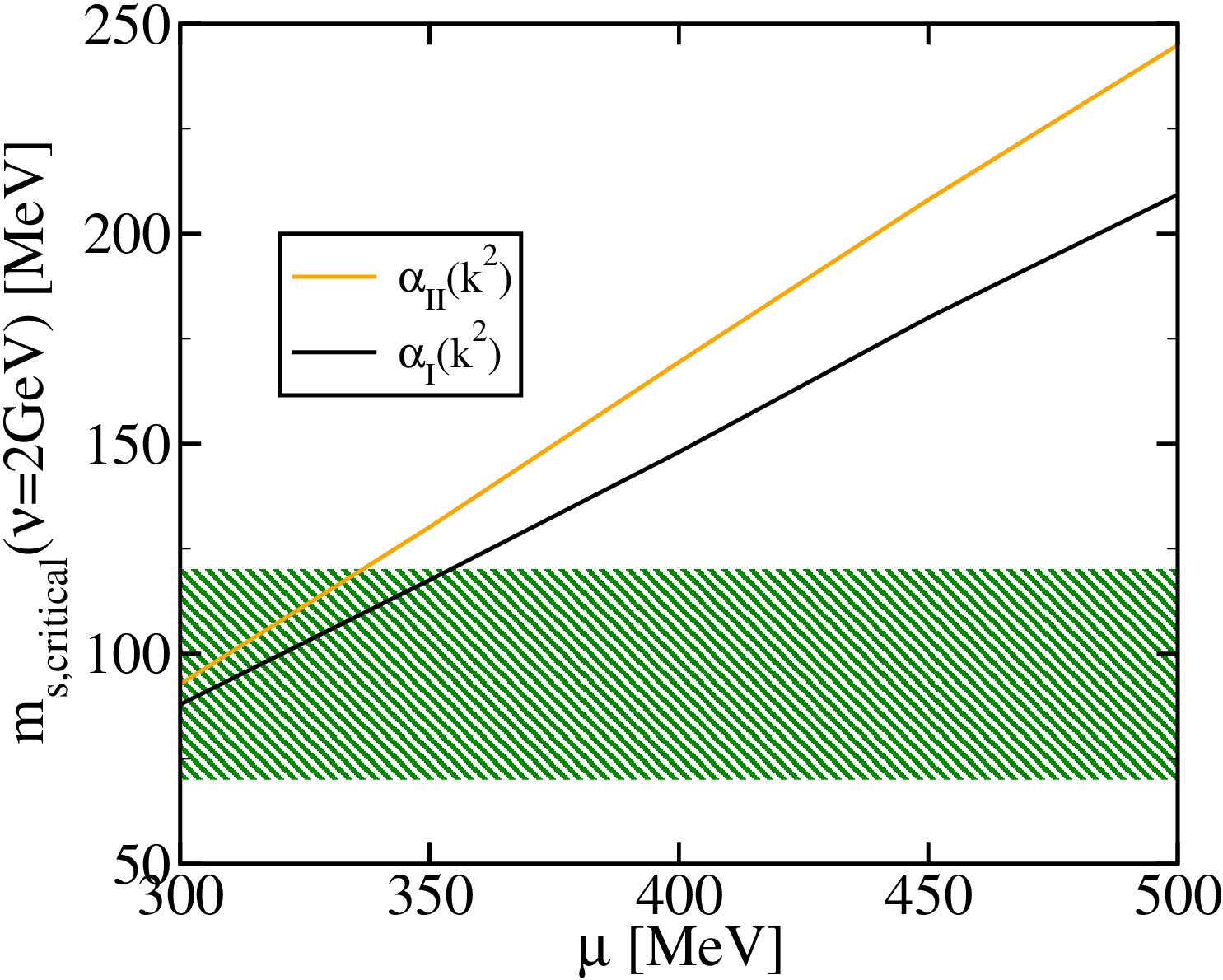}
  \end{center}
  \caption{Critical renormalized strange-quark mass as a function of the
    chemical potential for the couplings $\alpha_{I}(k^{2})$ and
    $\alpha_{II}(k^{2})$ and the range stated by the particle data
    group~\cite{PDG06} (shaded band). 
  }
  \label{mscritfig}
\end{figure}
In Fig.\ \ref{mscritfig} the results for the critical value of the renormalized
strange-quark mass as a function of the chemical potential are given and
compared to the range of the physical strange-quark current mass as determined 
by the particle data group~\cite{PDG06}.  (As stated above, the difference
between this mass in the $MOM$ and  the $\overline{MS}$ regularization scheme
is negligible compared to the experimental uncertainty.) As can be seen, the
approach taken here predicts that the physical strange-quark current mass is
very likely too small for allowing a 2SC phase at zero temperature for any
chemical potential. This result is remarkably stable against the variation of
the running coupling $\alpha_{s}(k^{2})$.

\section{Conclusions and outlook}
\label{conclusions}
We have studied the quark propagator in the 2SC and CFL phase at zero
temperature for different values of the strange-quark current mass in a fully
self-consistent Dyson-Schwinger approach to color-superconductivity.
Due to the medium modification of the interaction we find the 2SC phase to be
disfavored at any physically relevant chemical potential for the physical
value of the strange-quark current mass. This result is robust against
variation of the running of the strong coupling in the infrared within the given
uncertainties.
Since the CFL phase is only mildly influenced by neutrality conditions we
expect the result to hold even when neutrality is required. The critical
value of the strange-quark current mass should then be similar or even 
slightly larger. This will be explored in further studies.

Given the present results a further investigation of the CFL phase by the
inclusion of the Meissner effect and Goldstone-boson contributions is of
interest since both are expected to decrease the gap functions. Also the
temperature dependence and therefore the structure of the QCD phase diagram at
large densities is within the reach of the type of investigations presented
here. 

\section*{Acknowledgments}
We thank Michael Buballa and Dirk Rischke for helpful discussions, and Michael
Buballa for a critical reading of the manuscript.

D.N.\ is grateful to the DAAD for a fellowship, and to the members of the
FWF-funded  Doctoral Program ``Hadrons in vacuum, nuclei and stars'' at the
Institute of Physics of the University of Graz, where part of the work has
been done, for their hospitality.

In addition, this work has been furthermore supported in part by the Helmholtz
association (Virtual Theory Institute VH-VI-041) and by the BMBF under grant
number 06DA916.

\appendix*
\section{The color-flavor structure of the CFL phase}
We generalize the ansatz for the CFL phase with two degenerate quarks as given
in~\cite{Alford:1999pa} by choosing the matrices
\begin{eqnarray}
  P_{i} &=&
  \left(
    \begin{array}{ccccccccc}
      \delta_{i1}+\delta_{i2}&\delta_{i2}&\delta_{i4}&&&&&&\\
      \delta_{i2}&\delta_{i1}+\delta_{i2}&\delta_{i4}&&&&&&\\
      \delta_{i5}&\delta_{i5}&\delta_{i3}&&&&&&\\
      &&&\delta_{i1}&&&&&\\
      &&&&\delta_{i1}&&&&\\
      &&&&&\delta_{i7}&&&\\
      &&&&&&\delta_{i8}&&\\
      &&&&&&&\delta_{i7}&\\
      &&&&&&&&\delta_{i8}\\
    \end{array}
  \right),
  \\
  M_{i} &=&
  \left(
    \begin{array}{ccccccccc}
      \delta_{i1}+\delta_{i2}&\delta_{i2}&\delta_{i4}&&&&&&\\
      \delta_{i2}&\delta_{i1}+\delta_{i2}&\delta_{i4}&&&&&&\\
      \delta_{i5}&\delta_{i5}&\delta_{i3}&&&&&&\\
      &&&&\delta_{i1}&&&&\\
      &&&\delta_{i1}&&&&&\\
      &&&&&&\delta_{i8}&&\\
      &&&&&\delta_{i7}&&&\\
      &&&&&&&&\delta_{i8}\\
      &&&&&&&\delta_{i7}&\\
    \end{array}
  \right),
\end{eqnarray}
which fulfill the requirements of Eq.~(\ref{Urequire}). The basis is
defined by
\begin{eqnarray*}
  \{(r,u),(g,d),(b,s),(r,d),(g,u),(r,s),(b,u),(g,s),(b,d)\},
\end{eqnarray*}
with $r$, $g$, $b$ denoting the color and $u$, $d$, $s$ the flavor of the quarks.

\end{document}